\documentclass[aps,prd,amsmath,floats,floatfix, twocolumn, superscriptaddress,nofootinbib,showpacs]{revtex4-1}

\usepackage[colorlinks, pdfborder={0 0 0}, plainpages=false]{hyperref}
\usepackage{graphicx}
\usepackage{xspace}
\usepackage[usenames,dvipsnames]{color}
\usepackage{amssymb}
\usepackage{xcolor}
\usepackage{comment}
\usepackage[normalem]{ulem} 
\usepackage{array}
\usepackage{natbib}

\definecolor{mpl_red}{HTML}{D62728}

\newcommand\be{\begin{equation}}
\newcommand\ba{\begin{eqnarray}}
\newcommand\ee{\end{equation}}
\newcommand\ea{\end{eqnarray}}

\begin{document}

\title{A comprehensive look into the accuracy of spectral Einstein code binary black hole waveforms}

\author{Taylor Knapp}
\email{tknapp@caltech.edu}
\affiliation{TAPIR, California Institute of Technology, Pasadena, CA 91125, USA}
\affiliation{LIGO Laboratory, California Institute of Technology, Pasadena, California 91125, USA}

\author{Katerina Chatziioannou}
\email{kchatziioannou@caltech.edu}
\affiliation{TAPIR, California Institute of Technology, Pasadena, CA 91125, USA}
\affiliation{LIGO Laboratory, California Institute of Technology, Pasadena, California 91125, USA}

\author{Keefe Mitman}
\email{kem343@cornell.edu}
\affiliation{Cornell Center for Astrophysics and Planetary Science, Cornell University, Ithaca, New York 14853, USA}

\author{Mark A. Scheel}
\email{scheel@tapir.caltech.edu}
\affiliation{TAPIR, California Institute of Technology, Pasadena, CA 91125, USA}

\author{Michael Boyle}
\email{boyle@astro.cornell.edu}
\affiliation{Cornell Center for Astrophysics and Planetary Science, Cornell University, Ithaca, New York 14853, USA}

\author{Lawrence E. Kidder}
\email{kidder@astro.cornell.edu}
\affiliation{Cornell Center for Astrophysics and Planetary Science, Cornell University, Ithaca, New York 14853, USA}

\author{Harald Pfeiffer}
\email{harald.pfeiffer@aei.mpg.de}
\affiliation{Max Planck Institute for Gravitational
    Physics (Albert Einstein Institute), Am M\"{u}hlenberg 1, 14476
    Potsdam, Germany}

\date{\today}

\begin{abstract}
Numerical relativity simulations provide a full description of the dynamics of binary systems, including gravitational radiation. 
The waveforms produced by these simulations have a number of applications in gravitational-wave detection and inference. 
In this work, we revisit the accuracy of the waveforms produced by the Spectral Einstein Code. 
Motivated by the wide range of waveform applications, we propose and explore three accuracy metrics between simulation resolutions: (i) the generalized frequency-weighted mismatch, (ii) the relative amplitude difference, and (iii) the phase difference at different times. 
We confirm that numerical errors accumulate over the binary evolution, but the error is not intrinsically larger during the latest, more dynamical stages.
Studying errors across the parameter space, we identify a positive correlation between both the mismatch and the phase difference with precessing spin, but little correlation with aligned spin or eccentricity.  
Lastly, amplitude and phases differences are symmetric upon exchanging resolutions across the catalog, suggesting 
that the dominant source of error is random, rather than something systematic that affects all waveforms similarly.
\end{abstract}

\pacs{}

\maketitle

\section{Introduction} \label{sec:intro}

Numerical relativity (NR) simulations provide the most accurate and complete solution to the two-body problem in general relativity.
NR codes, such as the Spectral Einstein Code (\texttt{SpEC})~\cite{SXSCatalog}, have produced and will continue producing thousands of simulations for the coalescence of binary black hole (BBH) systems~\cite{Scheel:2025jct, ferguson2023secondmayacatalogbinary, Healy_2020, Rashti_2025, hamilton2023catalogueprecessingblackholebinarynumericalrelativity, Huerta_2019, Hinder:2013oqa, Aylott:2009tn}.
Though not the only one, a major application of these simulations is in gravitational-wave (GW) science, and interpreting the data of current~\cite{LIGOScientific:2014pky, VIRGO:2014yos} and future detectors~\cite{amaroseoane2017laserinterferometerspaceantenna, reitze2019cosmicexploreruscontribution, Branchesi:2023mws}.

Despite solving for the full dynamics exactly, BBH simulations are still subject to numerical errors.
Chief among them are those associated with the initial data \cite{Knapp:2024yww, Habib:2024soh, Ossokine_2015}, truncation error \cite{Boyle:2019kee, Szil_gyi_2014, Kopriva_2009}, and the GW extraction process \cite{Chu:2015kft, Chu_2016, Boyle:2009, Zlochower_2012}. 
Truncation error, in particular, grows during the transition from inspiral to merger, characterized by rapid dynamical changes.
Mitigating factors such as error-based time-stepping and adaptive mesh refinement~\cite{Boyle:2019kee, Lovelace_2011, Szil_gyi_2014} are designed to keep errors under control.
The overall simulation error in turn impacts the accuracy of the resulting waveforms and thus BBH data analysis.

Both the required \emph{level} of numerical accuracy and the \emph{metric with which we quantify it} depend on the eventual application.
Assuming convergence, numerical accuracy is typically proxied by an appropriately defined comparison between simulations of the same BBH system under different resolutions.
Concretely, comparisons rely on metrics such as the inner product (sometimes weighted by detector noise) between the waveforms~\cite{Buonanno_2007} as well as their phase difference~\cite{Zlochower_2012}.

The normalized noise weighted-inner product (hereafter referred to as the \emph{match}) is ideal for source inference applications and assessing accuracy across signal-to-noise ratios (SNRs)~\cite{jan2024accuracylimitationsexistingnumerical, Williamson_2017, Jan_2020, purrer2019readyliesahead}, as it is related to the likelihood~\cite{LIGOScientific:2019hgc}.
Previous work has established NR resolution thresholds for unbiased inference, including detections with future detectors~\cite{Ferguson_2021, Lousto_2023, purrer2019readyliesahead}.
For example, larger binary inclinations (more edge-on) and mass ratios (more unequal masses) generally require greater NR accuracy, due to higher harmonics, and thus higher resolutions~\cite{Ferguson_2021}. 
Furthermore, the match has revealed a quadratic scaling between the simulation length and the total error accumulated throughout the evolution~\cite{Mitman:2025tmj}.
Independently of detectors and sensitivity, NR waveforms are compared with approximate post-Newtonian waveforms to assess accuracy~\cite{Boyle_2007, Hinder_2010}, where discrepancies manifest as a dephasing that is more prevalent in noncircular and spin-precessing systems.

A key application of NR simulations is in the construction of waveform models which are used to analyze the data.
This includes direct surrogates of the full NR waveforms~\cite{Blackman_2015,Yoo_2022, Varma:2019, Varma_2019, Field_2014} or just the ringdown~\cite{MaganaZertuche:2024ajz, Pacilio:2024tdl,Nobili:2025ydt}, \footnote{Surrogates for the remnant properties have also been constructed and their accuracy has been quantified by the difference between the true and predicted quantity~\cite{Varma_2019, Thomas:2025rje}. In this study we restrict to waveform applications and do not consider the remnant properties further.
} and models that use NR waveforms in a more nuanced way~\cite{Pratten:2020ceb, Colleoni:2024knd, pompili_2023, Ramos-Buades:2023ehm, Chiaramello:2020ehz, Nagar:2020pcj, nagar_2023, Garcia-Quiros:2020qpx, Thompson:2023ase, Ghosh:2023mhc, Estelles:2025zah, Gamboa:2024hli, Boh__2017, Cotesta:2018fcv, Ossokine_2015, pompili_2023, Ramos-Buades:2023ehm, Nagar_2018, Gamba:2021ydi}. 
While the inspiral and ringdown stages for the latter models are motivated by the corresponding dynamical equations, the plunge and merger phases are calibrated to NR waveforms~\cite{Boh__2017, Ramos-Buades:2019uvh, Huerta:2016rwp, Hinder:2017sxy, Chattaraj:2022tay, Paul:2024ujx, pompili_2023, Gamboa:2024hli}.
The accuracy of the resulting model, be it a surrogate or phenomenological, is assessed against hold-out NR simulations that were not used in the model construction.
The main accuracy metric here is the match against NR waveforms.
In parameter estimation, mismatch thresholds are commonly defined to keep systematic parameter biases below statistical uncertainties, leading to an inverse scaling with signal-to-noise ratio~\cite{Lindblom:2008cm, Chatziioannou:2017tdw, thompson2025, Baird:2012cu, Owen:2023mid, Mezzasoma:2025moh}. 

Further applications target different portions of the signal, rather than the full signal.
They then test how well these portions can predict (or be consistent with) another portion of the signal~\cite{Bhat:2025lri, Tiwari:2025fua}.
One example is ringdown analyses based on NR ringdown surrogates~\cite{MaganaZertuche:2024ajz, Pacilio:2024tdl} or a set of quasinormal modes (QNM)~\cite{Isi:2019aib, Cotesta:2022pci, LIGOScientific:2025epi}.
While the ringdown spectrum is computed with perturbation theory~\cite{1973ApJ...185..635T, 1980ApJ...239..292D}, simulations are required for the initial QNM amplitude and phases~\cite{MaganaZertuche:2024ajz, Pacilio:2024tdl} and in general to study which modes are excited under what initial conditions~\cite{Nobili:2025ydt, Mitman:2025hgy}. 
Another example is tests of general relativity using different parts of the waveform~\cite{LIGOScientific:2021sio, LIGOScientific:2018dkp, LIGOScientific:2020tif, LIGOScientific:2016lio, LIGOScientific:2019fpa}, as no complete parametrized inspiral-merger-ringdown (IMR) model exists in alternative theories across parameter space and coupling constants~\cite{Lara:2025kzj, Okounkova:2019dfo, Okounkova:2019zjf, Okounkova:2020rqw}.
The more direct applications are the IMR-consistency test that splits the signal into two pieces according to their frequency, analyzes each independently, and then checks for consistency~\cite{Ghosh:2016qgn, Ghosh:2017gfp}, and tests of Hawking's area theorem that analyze the inspiral and postmerger phases independently~\cite{LIGOScientific:2025epi, Isi:2020tac}.
Further examples include tracing how information accumulates throughout the signal in frequency or time~\cite{Biscoveanu:2021nvg, Miller:2023ncs, Miller:2025eak} to (among other things) safeguard against data quality issues~\cite{Payne:2022spz, Udall:2024ovp}, as well as computations of the formation parameters of the binary in its distant past~\cite{Mould:2021xst, Varma:2021csh} or when it was potentially observable by a lower-frequency detector~\cite{Toubiana:2022vpp, Buscicchio:2024asl}. 

The above considerations call for numerical accuracy assessment metrics that go beyond the match, which is motivated by full-signal data analysis applications.
Since the match considers the whole waveform within specified integration limits and weights all frequencies equally (or according to the detector noise sensitivity), it is suboptimal for testing the accuracy of \emph{specific parts} of the waveform, which may be more or less relevant to parameter inference.
For example, waveform calibration and QNM studies motivate a metric that focuses on the merger phase and how accurately it matches the preceding inspiral.
The latter is also motivated by pure NR considerations as 
the underlying equations become increasingly nonlinear
at shorter timescales (i.e., merger) and numerical solutions increase in complexity. 
Numerical techniques such as adaptive mesh refinement (AMR) are designed to offset these challenges. 
Here we test in practice whether numerical errors grow toward the merger.

Additionally, the match is \emph{symmetric} upon exchanging the two waveforms compared. 
Therefore while it gives an estimate of the magnitude of the errors, it does not assess their sign.
This is relevant in waveform construction; even though the training and hold-out sets are distinct, they have been produced by the same code (or codes that employ similar techniques) and it is conceivable that numerical errors are correlated across the parameter space.
For example, if numerical dissipation causes all \texttt{SpEC} simulations to merge early, this bias would be incorporated into waveform models and would not be discernible by comparing to further simulations.
This bias can also be assessed through cross-code comparisons~\cite{Lovelace:2016uwp, Hinder:2013oqa, Aylott:2009ya, LIGOScientific:2014oec}.
However, the high computational cost and difficulty of generating long, high-accuracy simulations for the same system across multiple codes limit the scope of such studies. 
For this reason, we propose three metrics for waveform accuracy assessment utilizing SpEC simulations across different resolutions in Sec.~\ref{sec:methods}.
We assess waveform accuracy using multiple metrics applied to the largest and most extensive NR catalog currently available, thus conducting the most extensive accuracy study to date. 

The \emph{generalized mismatch} introduces a frequency weight into the inner product between the waveforms, extending the standard mismatch to probe frequency-dependent waveform accuracy.
In Sec.~\ref{sec:symmetric_analysis}, we explore weights that increasingly emphasize higher frequencies, thereby focusing on the late stages of the waveform.
Error accumulation over the course of numerical evolution has been long documented~\cite{Hannam:2007ik, Boyle_2007} and linked to waveform accuracy~\cite{Lindblom:2008cm, Zlochower:2012fk}. 
There have been efforts to assess accumulated phase error across NR codes~\cite{Hinder:2013oqa} and evaluate the accuracy of NR for future detectors where longer waveforms will be necessary~\cite{Wang:2024iyj}.
Reference~\cite{Mitman:2025tmj} demonstrated that waveform mismatch depends sensitively on the duration of the simulation, showing that comparisons based solely on mismatch can mischaracterize numerical accuracy for long NR waveforms.
We further propose the generalized mismatch as a means to mitigate the limitations of using the standard mismatch alone to characterize error accumulation in the waveform.
Our results are consistent with the finding in Ref.~\cite{Mitman:2025tmj}, namely that numerical errors accumulate throughout the evolution and are most pronounced near merger and ringdown.
However, considering the merger and ringdown in isolation, i.e., without the preceding inspiral, reveals that these stages remain highly accurate.
This suggests that \emph{the numerical solution does not lose accuracy as the dynamics become more complex, rather errors accumulate causing the late binary stages to appear less accurate}.
Furthermore, this confirms that \texttt{SpEC} can accurately evolve a system at all regimes and across the parameter space, regardless of the additional challenges of evolving Einstein's equations at shorter timescales (i.e., merger).
Across BBH parameters, we find that \emph{the generalized mismatch correlates only with the amount of precessing spin}, but not with the aligned spin or eccentricity. 

Switching to asymmetric accuracy metrics that capture the absolute error, we explore
\emph{relative waveform amplitude and phase differences} in Sec.~\ref{sec:asymmetric_analysis}.
These are computed per waveform mode as opposed to the full waveform.
We consider the dominant $(2,2)$ mode and subdominant $(2,1)$ and $(3,3)$ modes, which are most relevant for asymmetric systems, e.g., precessing spins or unequal masses binaries. 
There is \emph{no evidence of asymmetric errors in the waveform modes' amplitude and phases across the SXS catalog}. 
Roughly the same amount of simulations have positive and negative amplitude or phase differences between resolutions and for all modes.
However, like with the generalized mismatch, we find a \emph{positive correlation between the magnitude of the phase difference and the amount of precessing spin}.

This work systematically examines correlations between waveform accuracy and parameter space, especially asymmetric numerical errors as a function of BBH parameters, across an extensive catalog of NR waveforms.
Although our results are obtained with simulations of the \texttt{SpEC} code, we propose these metrics for a detailed assessment of the accuracy of other NR codes or waveform models.
We conclude in Sec.~\ref{sec:discussion}.

\section{Methods} \label{sec:methods}

In this section, we introduce three waveform accuracy metrics along with relevant technical details.
In Sec.~\ref{sec:waveform_alignment}, we establish notation and describe waveform conditioning. 
We recall the standard mismatch in Sec.~\ref{sec:mismatches} before introducing the generalized mismatch in Sec.~\ref{sec:nth_mismatches}. 
In Sec.~\ref{sec:asymmetric}, we define amplitude and phase differences for the asymmetric accuracy analysis.

\subsection{Waveform conditioning} \label{sec:waveform_alignment}

A waveform produced by \texttt{SpEC} for a set of physical parameters, $\lambda$, and observed from future null infinity is described in the inertial frame of the binary\footnote{
The inertial frame of the binary in SpEC is defined such that, at time $t=0$, the $z$-axis is aligned with the orbital angular momentum and the BHs lie along the $x$-axis, with the more massive BH on the positive $x$-axis.  
} as~\cite{Scheel:2025jct}
\begin{equation} \label{eqn:full_h}
    h(\lambda,\theta, \phi ;t) = \sum_{\ell, m} h^{(\ell,m)}(\lambda ;t)_{-2}Y_{(\ell, m)}(\theta, \phi)\,,
\end{equation}
where $_{-2}Y_{(\ell, m)}(\theta, \phi)$ are the spin-weighted $s=-2$ spherical harmonics for mode $(\ell, m)$ taken at angular position $(\theta, \phi)$ on the sky. 
In the source frame, $\theta$ is the angle between the line of sight and the direction of the binary orbital angular momentum at the start of the simulation (also known as the inclination $\iota$), and $\phi$ is the azimuthal angle~\cite{ajith2011dataformatsnumericalrelativity}.
The strain modes can be written as~\cite{Chu:2015kft, Varma:2019}
\begin{equation} \label{eqn:sph_harmonic_waveform}
    h^{(\ell,m)} (\lambda ;t) = A^{(\ell,m)} (\lambda ;t) e^{-i \Phi^{(\ell,m)} (\lambda ;t)}\,,
\end{equation}
where $A^{(\ell,m)} (\lambda ;t)$ and $\Phi^{(\ell,m)} (\lambda ;t)$ are the amplitude and phase for the $(\ell, m)$ mode as a function of time and physical parameters, $\lambda$.
The inertial reference frame of the simulation in which Eq.~\eqref{eqn:sph_harmonic_waveform} is defined with the BHs initially along the $x$-axis with the heavier-BH on the positive side and the $z$-axis parallel to the orbital angular momentum vector, $\vec{L}$. 

A BBH orbit is characterized by the mass ratio $q \equiv m_1/m_2 \geq 1$ between the primary mass $m_1$ and secondary mass $m_2$, eccentricity $e$, mean anomaly $l$, and the BH spins. 
The dimensionless spin vectors of the primary and secondary BHs $\vec{\chi}_1$ and $\vec{\chi}_2$ are specified in the inertial frame at the start of the simulation.
The total mass is an overall scaling and set to $M=1$ without loss of generality. 
These physical parameters, which we collectively denote as $\lambda$, enter the strain via the amplitudes and phases in Eq.~\eqref{eqn:sph_harmonic_waveform}. 
The latter are included in the NR data for each resolution of each simulation in the SXS catalog~\cite{Scheel:2025jct}.

We compare simulations of the same physical system at different numerical resolutions. 
Given initial conditions (that can be uniquely parametrized with $e$, $l$, $q$, $\vec{\chi}_1$, and $\vec{\chi}_2$ \cite{Knapp:2024yww}), \texttt{SpEC} solves Einstein's equations on discretized spacetime subdomains comprised of dense gridpoints~\cite{Boyle:2019kee}.
A higher resolution employs smaller spacetime subdomains with more densely populated gridpoints~\cite{Lovelace:2011, Szilagyi:2014} and thus is nominally more accurate.
Following common practice, we consider the two highest resolutions available for each simulation, denoting the highest and second highest resolutions by superscripts or subscripts ``I" and ``II" throughout, respectively.

\begin{figure}
    \centering
    \includegraphics[width=\columnwidth]{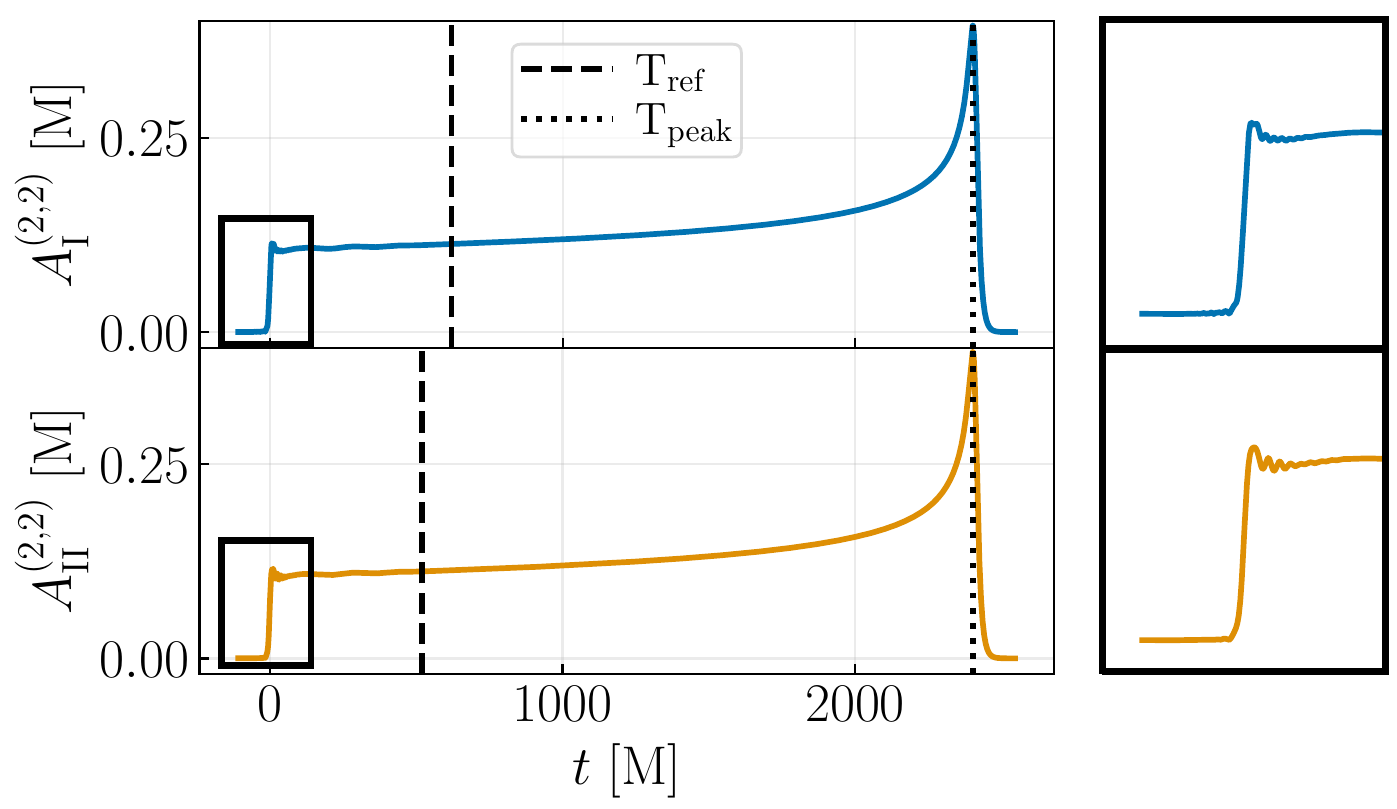}
    \caption{Waveform conditioning demonstrated with the amplitude of the $\ell=m=2$ mode $A^{(2,2)}_i$ for SXS:BBH:1141, which is a quasicircular, aligned-spin simulation with parameters given in Table \ref{tab:ex_sim_properties}. 
    The highest (second-highest) resolution is shown in the top (bottom) panel. Dashed vertical lines denote $T_{\rm ref}$, the approximate end of junk radiation for each resolution.
    The right-most plots enlarge the early portions of the waveform to highlight the junk radiation of each resolution (black boxes).
    Dotted vertical lines denote $T_{\rm peak}$, the time of largest amplitude for each resolution. Both times differ between resolutions.
    When comparing two resolutions, we pick the latest of the two $T_{\rm ref}$'s and the $T_{\rm peak}$ of the highest resolution.
    }
    \label{fig:alignment_h_1979}
\end{figure}

After calculating Eq.~\eqref{eqn:full_h} from the raw catalog contents, a number of conditioning steps are required. 
First, waveforms contain a phase of transient ``junk radiation," which refers to the gravitational radiation emitted as the simulation relaxes to quasiequilibrium.
The end of junk radiation can be determined via an adaptive algorithm~\cite{Pretto:2024dvx} or more crudely by tracking changes to the average irreducible mass of the simulation \cite{Boyle:2019kee, Higginbotham:2019wbx}, which systematically produces an underestimate of the end of junk radiation.
Among various methods that have been proposed to determine the end of junk radiation ~\cite{Husa:2015iqa, Hamilton:2021pkf}, the current catalog uses the adaptive algorithm from ~\cite{Pretto:2024dvx} to determine a time $T_{\rm ref}$ after which junk radiation has exited the domain~\cite{Knapp:2024yww, Habib:2024soh}, which in general is different for each resolution. 
Figure~\ref{fig:alignment_h_1979} shows an example.
Junk radiation is subtly different for each resolution since different densities of grid spacing relax into different equilibriums and adaptive grid spacing is not always implemented until the end of junk radiation.
In the example in Fig.~\ref{fig:alignment_h_1979}, although the junk radiation appears differently between the resolutions, it is most pronounced around $t=0$ for both.
As a consequence, $T_{\rm ref}$ (dashed vertical line) differs.
In order to obtain waveforms of comparable length, we adopt the later of the two $T_{\rm ref}$'s and discard the data before that time for both waveforms. 

Second, the waveforms $h_I$ and $h_{II}$ must be appropriately aligned.
Waveforms at future null infinity possess a large number of symmetries: supertranslations, rotations, and boosts that we can choose to optimize over when comparing two waveforms~\cite{Mitman:2021xkq, DaRe:2025glj, Boyle:2015nqa}. 
Following Ref.~\cite{Boyle:2019kee}, we optimize over time, $t$, and azimuthal angle, $\phi$, 
via a time shift $\delta t$ and a phase shift $\delta \phi$\footnote{The parameter $\phi$ in Eq.~\eqref{eqn:full_h} corresponds to the azimuthal angle and $\delta \phi$ to a shift of this angle. In other contexts, $\phi$ may refer to the phase of the waveform. The two are not equivalent in the presence of spin-precession and higher-order modes~\cite{Miller:2025eak}.  
}
in Eq.~\eqref{eqn:full_h}
\begin{equation} \label{eqn:full_h_shift}
    h(t) \rightarrow \sum_{\ell, m} h^{(\ell,m)}(t + \delta t)_{-2}Y_{(\ell, m)}(\theta, \phi) e^{im \delta \phi}\,.
\end{equation}  
Different choices for the time shift $\delta t$ and the phase shift $\delta \phi$ are more appropriate for different metrics (generalized mismatch, amplitude ratio, phase differences) and applications.\footnote{Additionally aligning over inclination, $\theta$, would correspond to further optimization. 
We choose not to do so because the inclination carries a clear physical meaning and measuring it is desirable.}
We will return to these in subsequent subsections alongside their pertinent accuracy metrics.

\begin{figure}
    \centering
    \includegraphics[width=\columnwidth]{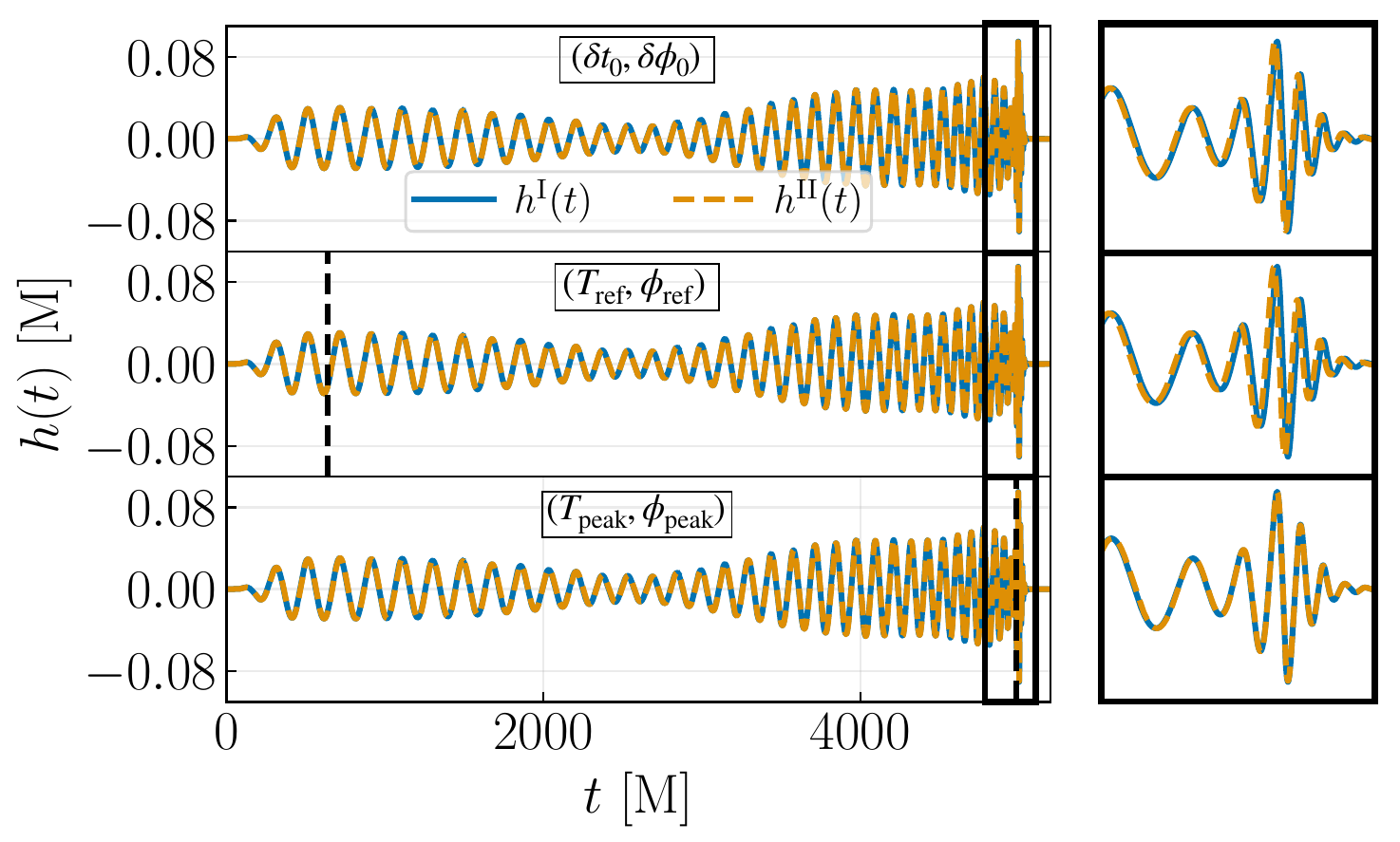}
    \caption{Demonstration of various time and phase alignment schemes used throughout this work between the highest (blue) and second-highest (orange) resolutions of SXS:BBH:0803. 
    The left column shows the full waveform while the right column enlarges the merger portion (black boxes). 
    In the top row, we align waveforms by $(\delta t_0, \delta \phi_0)$, obtained by minimizing the normalized $L^2$ norm between resolutions, see Sec.~\ref{sec:mismatches}.
    In the second row, we align the waveforms at the beginning of the usable part of the waveforms, $(T_{\rm ref},\phi_{\rm ref})$  (dashed vertical line). 
    In the third row, we align them at the peak, $(T_{\rm peak},\phi_{\rm peak})$ (dashed vertical line).
    We refer to Sec.~\ref{sec:asymmetric} for discussion about the alignments for the second and third rows.
    Each alignment choice trades the amount of (dis)agreement between different parts of the waveform. 
    \label{fig:alignment_ex_1979}}
\end{figure}

\subsection{Mismatch} \label{sec:mismatches}

Given $h^{\rm I}$ (highest resolution waveform), its agreement with an appropriately time- and phase-shifted $h^{\rm II}$ (second-highest resolution) is quantified through the \textit{overlap} \cite{Buonanno_2007,LIGOScientific:2019hgc}. Here we adopt the generalization~\cite{Boyle:2019kee}
\begin{equation} \label{eqn:overlap}
    \mathcal{O} (h^{\rm I}, h^{\rm II}; \delta \phi, \delta t) = \frac{( h^{\rm I} | h^{\rm II} (\delta \phi, \delta t) )}{\sqrt{( h^{\rm I} | h^{\rm I} ) ( h^{\rm II} (\delta \phi, \delta t)| h^{\rm II} (\delta \phi, \delta t) )}}  \,,
\end{equation}
where the sky-averaged, noise-weighted inner product is
\begin{equation} \label{eqn:inner_product}
    ( a|b ) = 4 {\rm Re}\int\int_{S^2} \frac{\tilde{a}(f)\tilde{b}^*(f)}{S_n(|f|)}d\Omega df\,.
\end{equation}
This is integrated over frequency over the whole simulation waveform and the two-sphere, ${\rm Re}()$ is the real operator, tilde denotes a Fourier transform, and $S_n(|f|)$ is the one-sided power spectral density (PSD) that encodes the detector's sensitivity.  
We adopt a flat PSD of $S_n(|f|) = 1$, thus avoiding any additional frequency weighting before we introduce our own in Sec.~\ref{sec:nth_mismatches}. 
The time shift, $\delta t$, is a real shift of $h^{\rm II}$ in time.
The phase shift, $\delta \phi$, is an azimuthal angular rotation of the system that changes the location of the observer with respect to the source \cite{Boyle:2019kee}.
This shift only changes the angle at which the observer views the system, not any of the underlying physics of the binary.

Although $\delta t$ and $\delta \phi$ are commonly determined by maximizing Eq.~\eqref{eqn:overlap}, we follow Ref.~\cite{Boyle:2019kee} that proposes an alternative, described in Appendix~\ref{apx:L2_minimization}.
Since Eq.~\eqref{eqn:overlap} is normalized and thus invariant to an overall amplitude difference between the waveforms, we obtain $\delta t$ and $\delta \phi$ by instead minimizing the normalized $L^2$ norm, 
\begin{alignat}{2} \label{eq:L2_norm}
    \hat{L}^2 &= \frac{( h^{\rm I} - h^{\rm II} | h^{\rm I} - h^{\rm II} )}{( h^{\rm I} | h^{\rm I})}\,. 
\end{alignat}
Since $\hat{L}^2$ is not normalized by \emph{both} waveforms, minimizing $\hat{L}^2$ achieves the same effect as maximizing the overlap while also accounting for scaling differences between $h^{\rm I}$ and $h^{\rm II}$. 
The top row of Fig.~\ref{fig:alignment_ex_1979} gives an example of this alignment procedure for a precessing waveform.
Alignment is dominated by portions of the waveform that are ``louder'' in the frequency domain, i.e., the inspiral, as this is where the largest contribution to the $\hat{L}^2$ norm comes from and differences have a greater impact.

Finally, as waveforms are very similar to each other and overlaps cluster around unity, we present the \emph{mismatch}
\begin{equation} \label{eqn:mismatch}
   \overline{ \mathcal{M}}(h^{\rm I}, h^{\rm II}, \delta \phi, \delta t) = 1 - \mathcal{O} (h^{\rm I}, h^{\rm II}, \delta \phi, \delta t)\,.
\end{equation}
Since both the overlap definition and the minimization include an integral over the two-sphere, the final quantity corresponds in some sense to the \textit{averaged mismatch} instead of the mismatch used for data analysis.
Following the convention in Ref.~\cite{Scheel:2025jct}, we denote it with a bar but refer to it as a ``mismatch" throughout. 

Both the functional form of the overlap and the alignment methodology are motivated by and related to matched-filtering techniques for GW data analysis. 
The expectation value of the log-likelihood for a signal with parameters $\lambda$ in data \textbf{d}, is given by Eq.~(30) in Ref.~\cite{LIGOScientific:2019hgc}
\begin{equation} \label{eqn:exp_val_likelihood}
    {\rm E} [{\rm log} \Lambda (\textbf{d} | \lambda)] = \frac{1}{2} \mathcal{O}^2 (\textbf{d}, \lambda) {\rm SNR}^2\,.
\end{equation}
Here the overlap is between the data and the filter, i.e., a waveform with parameters $\lambda$. 
The overlap is calculated in its noise-weighted form with $S_n(|f|)$ corresponding to the detector sensitivity and upweighting frequencies for which the detector noise is lower.
In order to achieve a high likelihood and identify a signal, we need a loud signal and a filter that matches the data as quantified by the overlap.
A similar formula links the overlap between the true signal and an approximate waveform with which we model it to the systematic bias in inference~\cite{Chatziioannou:2017tdw}
\begin{equation} \label{eq:overlap_ineq}
    1 - \mathcal{O}(\textbf{d}, \lambda) < \frac{D}{2{\rm SNR}^2}\,,
\end{equation}
where $D$ is the relevant number of model parameters in $\lambda$. 
This inequality indicates that higher-SNR signals require smaller mismatches between the true and analysis waveforms, so that statistical uncertainties from finite SNR remain smaller than systematic errors from model inaccuracies.

\subsection{Generalized mismatch} \label{sec:nth_mismatches}

The mismatch defined in Eq.~\eqref{eqn:mismatch} adopts a weighting of the waveform frequencies
according to the PSD, which encodes the detector's sensitivity.
With this definition, targeting different portions of the waveform is typically pursued by varying the binary total mass. 
However, this effective ``frequency weighting" does not allow full flexibility in choosing which portion of the waveform to target.
This is because realistic noise curves are complex functions with shapes determined by the detector rather than waveform-related considerations,
typically approximating a power law up to about $100$\,Hz before flattening and rising again.
This rigidity makes it difficult to target specific portions of the waveform when computing the mismatch.

To increase flexibility, we define the \textit{generalized mismatch} via a general form of the inner product that allows for a user-specified frequency-weighting function
\begin{equation}
    \label{eqn:inner_product_Ff}
    ( h^{\rm I} | h^{\rm II} ) = 4 {\rm Re}  \int \int _{S^2} \tilde{h}^{\rm I} (f) \tilde{h}^{\rm II,*} (f) \mathcal F(f) d\Omega  df\,. 
\end{equation}
Here $\mathcal F(f)$ is a general frequency weighting that is not restricted to a detector-sensitivity PSD and can be tailored to specific portions of the waveform, providing a mismatch with adjustable frequency weighting. 

In what follows, we are motivated by PN calculations that suggest that the waveform's amplitude follows roughly a power-law frequency evolution toward merger. 
We therefore choose $\mathcal F_n(f) = f^n$ in Eq.~\eqref{eqn:inner_product_Ff}, where $n$ is a user-tunable parameter for emphasizing different portions of the waveform

\begin{equation} \label{eqn:inner_product_weighted}
    ( h^{\rm I} | h^{\rm II} ) _n = 4 {\rm Re}  \int \int _{S^2} \tilde{h}^{\rm I} (f) \tilde{h}^{\rm II,*} (f) f^n d\Omega  df  \,,
\end{equation}
and obtain the frequency-weighted overlap and mismatch by analogy to Eqs.~\eqref{eqn:overlap} and~\eqref{eqn:mismatch}, see Appendix \ref{apx:L2_minimization}, and denote it with a subscript of $n$ to distinguish it from the standard mismatch.
The choice $n=0$ reduces to the standard mismatch,  $n > 0$ upweights higher frequencies (late inspiral and merger), while $n <0$ upweights lower frequencies (early inspiral).

To visualize the weighted waveforms, we compute the frequency-weighted strain in the time domain~\cite{boyle2025sxs}: 
\begin{equation} \label{eqn:n_deg_strain}
    h_{n}(t) = \int ^\infty _{-\infty} \tilde{h}(f)f^{\frac{n}{2}} e^{i 2\pi ft} df\,.
\end{equation}
For $n=0$, we recover the original waveform.
We consider up to $n=3$, which effectively produces inspiral-less waveforms.
In the frequency domain, the inspiral scales as $\tilde{h}(f)\sim f^{-7/6}$.
For $n=3$, therefore, the weighted waveform scales as $\tilde{h}(f)f^{3/2}\sim f^{1/3}$.
This transition from negative to positive exponents of $f$ signifies the shift from inspiral-dominated to merger-dominated results.

\begin{figure}
    \centering
    \includegraphics[width=\columnwidth]{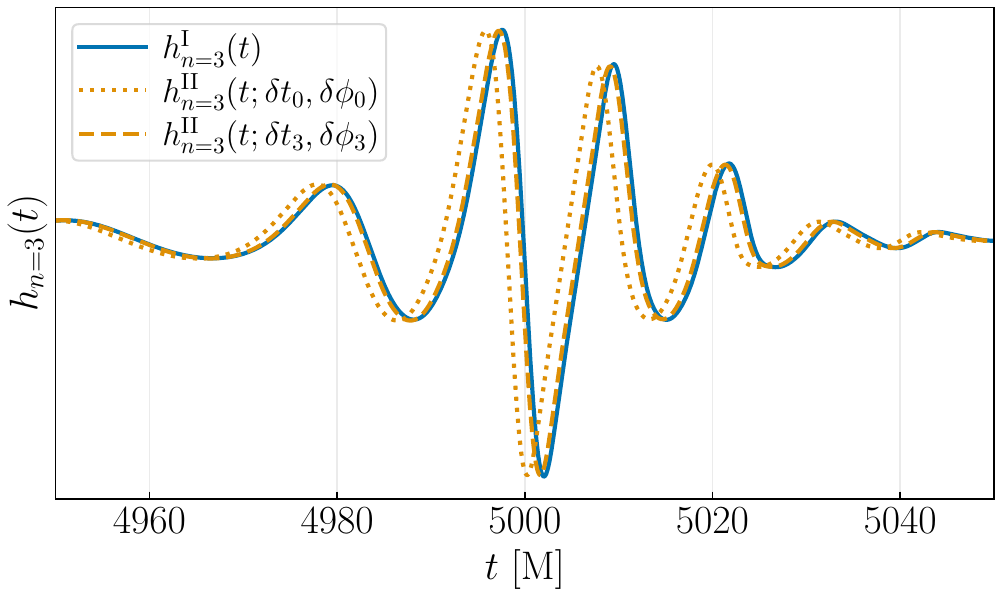}
    \caption{ 
    Comparison of alignment prescriptions for SXS:BBH:0803 for $n=3$.
    The highest resolution, $h_{n=3}^{\rm I}(t)$, is plotted in blue. 
    We also plot the second highest resolution, $h_{n=3}^{\rm II}(t)$, shifted by $(\delta t_0,\delta \phi_0)$ (orange dotted) and by $(\delta t_3,\delta \phi_3)$ (orange dashed). 
    When realigning after frequency-weighting the waveform, the two resolutions show better agreement at merger (blue and orange dashed) than when with the one-time alignment (blue and orange dotted).
    }
    \label{fig:align_compare_merger}
\end{figure}

The generalized mismatch, $\overline{ \mathcal{M}}_{n,m}$, follows from Eq.~\eqref{eqn:mismatch} by analogy but has the flexibility to use generic time and phase shifts: 
\begin{alignat}{2} \label{eqn:mismatch_weighted}
    \overline{ \mathcal{M}}_{n,m}(&h^{\rm I}, h^{\rm II},\delta \phi_m, \delta t_m) \nonumber\\
        &= 1 - \frac{( h^{\rm I} | h^{\rm II} (\delta \phi_m, \delta t_m) )_n}{\sqrt{( h^{\rm I} | h^{\rm I} )_n ( h^{\rm II} (\delta \phi_m, \delta t_m)| h^{\rm II} (\delta \phi_m, \delta t_m) )_n}}\,,
\end{alignat}
where the subscript $n$ denotes the order of the frequency weighting and the subscript $m$ denotes that the shifts are obtained from minimizing the $L^2$ norm between the frequency-weighted waveforms $h_m$.
Below we adopt two alignment prescriptions:
\begin{enumerate}
    \item Fixed $(\delta t_0, \delta \phi_0)$ for all mismatches. 
    We compute $(\delta t_0, \delta \phi_0)$ from the unweighted waveforms (equivalently, $n=0$) and then use them to compute all weighted mismatches, denoted $\overline{ \mathcal{M}}_{n,0}$. This prescription compares the waveform mergers in relation to their preceding inspirals.
    \item Minimize over $(\delta t_{m=n}, \delta \phi_{m=n})$ for each $n$.  
    We compute $(\delta t_n, \delta \phi_n)$ from the $n$-weighted waveforms and then use them to compute the corresponding weighted mismatch, denoted $\overline{ \mathcal{M}}_{n,n}$. This prescription compares the waveform mergers in isolation, with no impact from the preceding inspiral.
\end{enumerate}

Figure~\ref{fig:align_compare_merger} illustrates these prescriptions for the second highest resolution of SXS:BBH:0803 and $n=3$, compared to the highest resolution.
As expected, the second prescription leads to a better visual agreement at merger.
In Sec.~\ref{sec:symmetric_analysis}, we apply both prescriptions on the SXS catalog.


\subsection{Asymmetric accuracy metrics} \label{sec:asymmetric}

The mismatch has two disadvantages that motivate further metrics.
First, it is symmetric with respect to the waveforms: $\overline{ \mathcal{M}}_{n,m}(h^{\rm I}, h^{\rm II}) = \overline{ \mathcal{M}}_{n,m}(h^{\rm II}, h^{\rm I})$.
Therefore while it captures the magnitude of their difference, it does not capture its sign.
Second, since it is normalized, it is insensitive to a constant amplitude difference between the waveforms.
We propose and study two asymmetric quantities: the relative amplitude difference and phase difference of each GW spherical harmonic mode from Eq.~\eqref{eqn:sph_harmonic_waveform} defined respectively as
\begin{align} \label{eq:asymm_quantities_R}
    \Delta A^{(\ell,m)}(t) &= \frac{A^{(\ell,m)}_{\rm I}(t) - A^{(\ell,m)}_{\rm II}(t)}{\sqrt{A^{(\ell,m)}_{\rm I}(t)\times A^{(\ell,m)}_{\rm II}(t)}} \,,\\
    \Delta \Phi^{(\ell,m)} (t) &= \Phi^{ (\ell,m)}_{\rm I} (t) - \Phi^{(\ell,m)}_{\rm II}(t)\,.\label{eqn:phase_diffs}
\end{align}

Both are functions of time and encode the full amplitude and phase evolution throughout the binary coalescence.
Perfectly matching waveforms yield $0$ for both quantities.
Furthermore, both quantities reveal asymmetries between the simulation resolutions.
A positive $\Delta A^{(\ell,m)}$ or $\Delta \Phi^{(\ell,m)}$ implies that the higher resolution has a larger amplitude or phase respectively.
A consistent sign for either $\Delta A^{(\ell,m)}$ or $\Delta \Phi^{(\ell,m)}$ across the entire catalog indicates the presence of systematic errors in SpEC that would not be evident in the mismatch distribution.
For example, if the majority of $\Delta A^{(\ell, m)}$ are positive, we could conclude that higher resolutions produce waveforms with systematically greater amplitudes than lower resolutions.
On the other hand, if $\Delta \Phi^{(\ell,m)}$ had a sign across the entire catalog, that could be indicative of numerical dissipation. 
Such errors would also be calibrated into analytical models and could remain indiscernible from mismatch comparisons against noncalibration simulations.

While a constant amplitude difference leaves the mismatch unchanged, the phase difference is already probed by the mismatch.
Considering it here as a function of time rather than integrated over frequency allows us to explore where exactly errors arise throughout the coalescence.
Moreover, the mode amplitudes are more strongly affected by finite differencing, extraction, and gauge effects~\cite{Boyle:2015nqa, Boyle:2009}, while phase differences are largely driven by truncation error.
We therefore expect the former to be larger and indicative of systematic errors.

Evaluating Eqs.~\eqref{eq:asymm_quantities_R}-\eqref{eqn:phase_diffs} again requires some time and phase alignment. 
We consider three prescriptions, depicted in Fig.~\ref{fig:alignment_ex_1979}:
\begin{enumerate}
    \item Minimize over $(\delta t_0, \delta \phi_0)$ (top row). This is the same alignment obtained by minimizing $\hat{L^2}$ in Eq.~\eqref{eq:L2_norm}, and it is the most closely related prescription to GW data analysis for the full signal.

    \item Align via $(T_{\rm ref},\phi_{\rm ref})$ (second row). Here, $t=T_{\rm ref}$ marks the beginning of the postjunk waveforms.
    To align the phases, we begin to unwrap each GW phase at $t=T_{\rm ref}$ so that $\Delta\Phi^{(\ell, m)}(t=T_{\rm ref}) = 0$.
    This prescription targets the question of how well we can predict the later parts of the waveform given the inspiral.

    \item Align at $(T_{\rm peak},\phi_{\rm peak})$ (third row).
    We adopt the same procedure as above but for $T_{\rm peak}$, the peak of the strain amplitude.
    This prescription targets the question of how well we can predict the early parts of the waveform given the merger.
\end{enumerate}

\section{Results: Mismatches} \label{sec:symmetric_analysis}

In this section, we calculate the generalized mismatch outlined in Sec.~\ref{sec:nth_mismatches} across the SXS catalog~\cite{Scheel:2025jct}.
We first discuss three select simulations in detail in Sec.~\ref{sec:mismatch_ex}, before tackling all BBH simulations in Sec.~\ref{sec:mismatches_full}.  

\subsection{Example simulations} \label{sec:mismatch_ex}

\renewcommand{\arraystretch}{1.25}
\begin{table}
    \centering
    \begin{tabular}{|c|c|c|c|c|}
        \hline
        Simulation & $e$  & $l$ & $\vec{\chi}_1(t=T_{\rm ref})$ & $\vec{\chi}_2(t=T_{\rm ref})$ \\ \hline
        SXS:BBH:1141 & 0  & \textcolor{red}{$\cdots$} & (0.00,0.00,-0.44) & (0.00,0.00,0.44)\\ \hline
        SXS:BBH:1356 & 0.16&0.79& (0.00,0.00,0.00) & (0.00,0.00,0.00) \\ \hline
        SXS:BBH:0803 & 0  & \textcolor{red}{$\cdots$} & (0.61,0.34,-0.40)& (0.77,0.01,0.22)\\ \hline
    \end{tabular}
    \caption{Parameters for each example simulation of Sec.~\ref{sec:mismatch_ex}. We list the eccentricity, mean anomaly, and dimensionless spin vectors at $t=T_{\rm ref}$ in the inertial frame of the binary. All simulations have a mass ratio of $q=1$.
    }
    \label{tab:ex_sim_properties}
\end{table}

\begin{figure*}
    \centering
    \includegraphics[width=\textwidth]{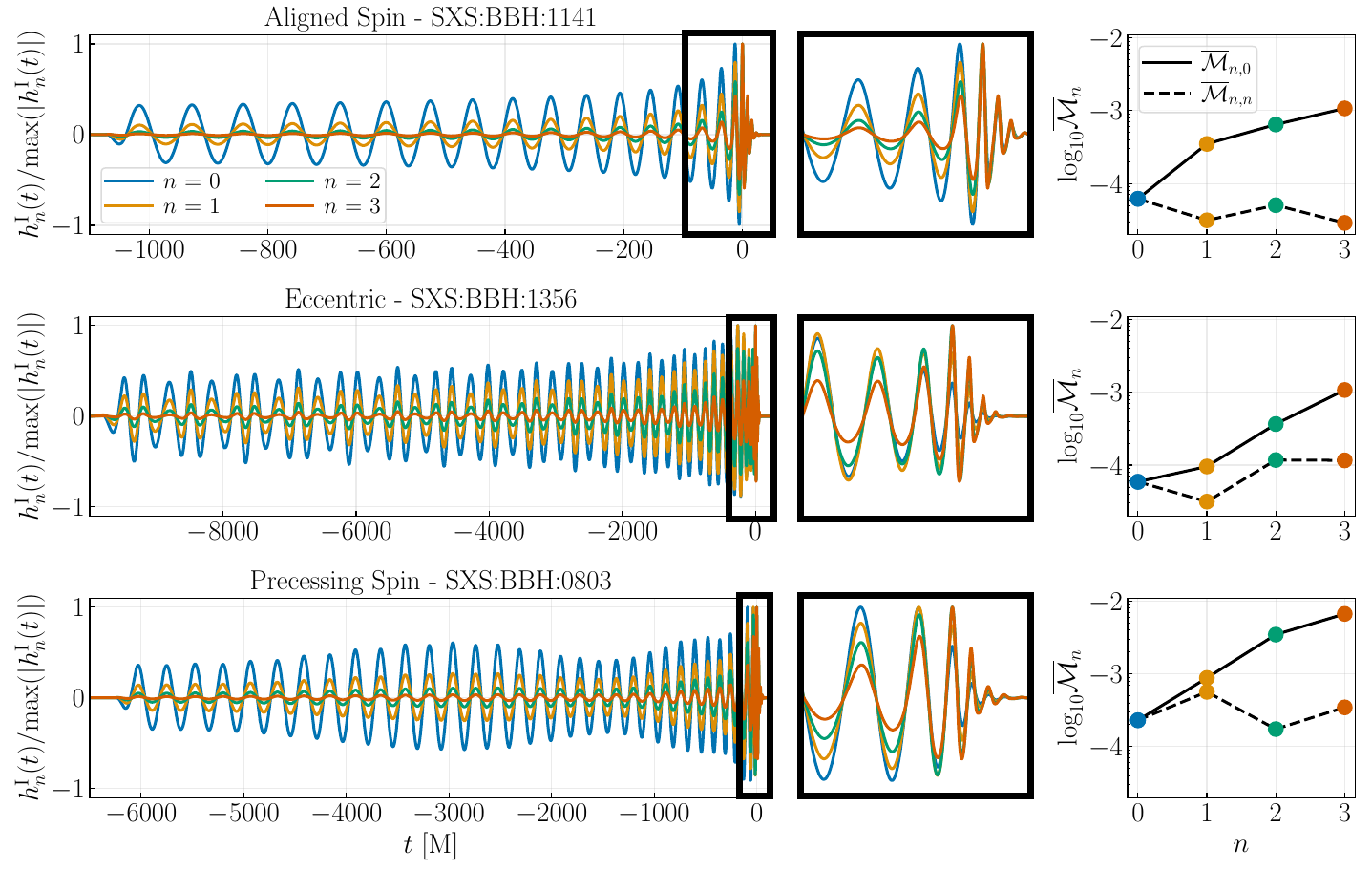}
    \caption{Normalized weighted waveforms and mismatches for three example simulations: the quasicircular spin-aligned SXS:BBH:1141 (top row), the nonspinning eccentric SXS:BBH:1356 (middle row), and the quasicircular spin-precessing SXS:BBH:0803 (bottom).
    Parameters for each are given in Table~\ref{tab:ex_sim_properties}. 
    We plot the frequency-weighted strain up to $n=3$, Eq.~\eqref{eqn:n_deg_strain}, for the highest resolution in the left column. Strains are normalized by the peak amplitude, ${\rm max}|h_n(t)|$. 
    Since the mismatch is scale-invariant, this normalization does not affect the mismatch and the visual comparison is faithful.
    The merger portion of each weighted waveform is shown in the middle column (black box). 
    Generalized mismatches for two alignments, $\overline{ \mathcal{M}}_{n,0}$ (solid lines) and $\overline{ \mathcal{M}}_{n,n}$ (dashed lines), are shown in the right column.
    As $n$ increases, the inspiral is downweighted compared to the merger, so generalized mismatches for larger $n$ are increasingly targeting the later coalescence stages.}
    \label{fig:mismatches_ex}
\end{figure*}

\begin{figure*}
    \centering
    \includegraphics[width=\textwidth]{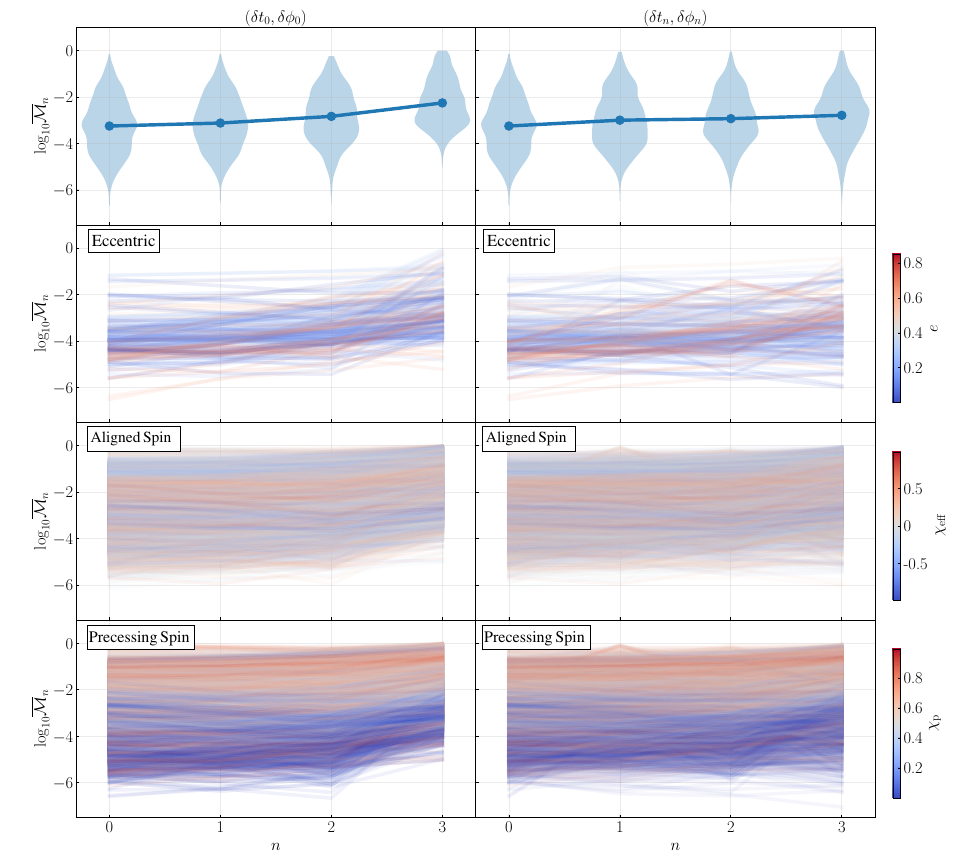}
    \caption{Generalized mismatches $\overline{ \mathcal{M}}_{n,0}$ (left column) and $\overline{ \mathcal{M}}_{n,n}$ (right column) as a function of the degree of frequency weighting $n$ for all BBH simulations in the SXS catalog. 
    The top row shows distributions for all simulations with the solid lines connecting the means. 
    The bottom three rows show results for individual simulations (lines) separated into three categories and colored by pertinent parameters: eccentric (second row; colored by $e$), aligned spin (third row; colored by $\chi_{\rm eff}$), and precessing (bottom row; colored by $\chi_p$) simulations. 
    The accuracy of precessing simulations is anticorrelated with $\chi_p$. 
    \label{fig:mismatches_all}}
\end{figure*}

\begin{figure*}
    \centering
    \includegraphics[width=\textwidth]{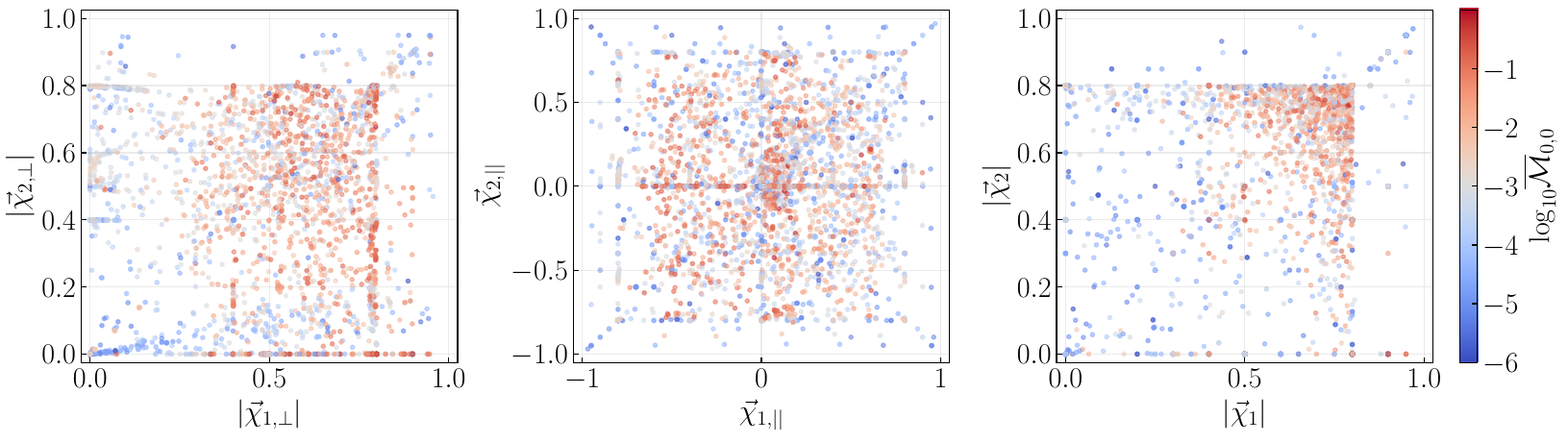}
    \caption{Scatter plots showing the log standard mismatch, $\log_{10}\overline{\mathcal M}_{0,0}$, for each simulation in the catalog (color bar) as a function of different spin components. 
    We present the in-plane spin magnitude (left), aligned spin component (middle), and total spin magnitudes (right). 
    Greater total spin and in-plane spin magnitudes tend to have larger mismatches.
    \label{fig:scatter_mismatches}}
\end{figure*}

We begin with an in-depth discussion of three representative BBH simulations exemplifying systems with aligned spins, spin-precession, and eccentricity, respectively.
The parameters for each simulation are listed in Table~\ref{tab:ex_sim_properties}. 
Weighted waveforms and generalized mismatches are shown in Fig.~\ref{fig:mismatches_ex}.
For each simulation (top to bottom), we plot the weighted strain $h_n(t)$ and the mismatch under both alignment prescriptions described in Sec.~\ref{sec:nth_mismatches}.
As the value of $n$ increases, the weighted waveforms become increasingly merger-dominated.
The inspiral is progressively downweighted, while the merger and ringdown are upweighted.
In the aligned-spin case, the amplitude envelope rises monotonically, and frequency weighting makes this increase more pronounced.
In the precessing and eccentric cases, the amplitude envelope exhibits modulations on different timescales.
As these modulations are of lower frequency than the orbital frequency, they are also downweighted.
As a result, by $n=3$ the waveforms have nearly monotonic amplitude envelopes.
In all cases, frequency weighting smooths out and removes the inspiral, effectively targeting the merger without artificially terminating the waveform in time or frequency.

We now turn to the generalized mismatches.
All simulations are highly accurate with a standard mismatch of $\overline{ \mathcal{M}}_{0,0}{\sim} 10^{-4}$.
Keeping the alignment $(\delta t_0, \delta \phi_0)$ determined by the full waveform, we then progressively upweight the merger with $n>0$.
The resulting generalized mismatches $\overline{ \mathcal{M}}_{n,0}$ monotonically increase as a function of $n$, reaching values of ${\sim}10^{-3}$ for the two nonprecessing and ${\sim}10^{-2}$ for the precessing simulations respectively.
The inclusion of the long inspiral in determining $(\delta t_0, \delta \phi_0)$ results in $\overline{ \mathcal{M}}_{n,0}$ being dominated by ensuring alignment between resolutions during the inspiral phase. 
This alignment prescription is therefore increasingly suboptimal for the later waveform stages as revealed by the increasing values of $\overline{ \mathcal{M}}_{n,0}$.
We confirm that numerical error is monotonically accumulating throughout the coalescence, and the merger is increasingly inconsistent between resolutions compared to the inspiral~\cite{Mitman:2025tmj}.

We next target a more nontrivial question: is simulating merging BHs inherently more inaccurate than inspiraling BHs due to the complexity of the equations that are numerically evolved?
We address this question with $\overline{\mathcal{M}}_{n,n}$, the generalized mismatch where waveforms are optimally aligned \emph{after} frequency weighting.
We find that now $\overline{ \mathcal{M}}_{n,n}$ stays roughly constant for all simulations around ${\sim}10^{-4}$ and strictly $\overline{ \mathcal{M}}_{n,n}<\overline{ \mathcal{M}}_{0,0}$.
The latter is expected due to the additional optimization in $\overline{ \mathcal{M}}_{n,n}$ compared to $\overline{ \mathcal{M}}_{0,0}$.
The former suggests that $\overline{ \mathcal{M}}_{0,0}$ increases \emph{due to suboptimal inspiral-dominated alignment and not because the merger is inherently less accurate than the inspiral}.

\subsection{Full catalog} \label{sec:mismatches_full}

Having gained some intuition about the properties of generalized mismatches and the role of alignment, we turn to the full SXS catalog~\cite{Scheel:2025jct}.
Of the total 3,756 vacuum BBH simulations, we use the catalog metadata to select the 3,605 simulations that have at least two resolutions, have not been deprecated, and have been run with a robust version of \texttt{SpEC} (see Sec.~3.1 of Ref.~\cite{Scheel:2025jct} for details). 
In practice, this leaves us mainly with simulations after ID SXS:BBH:0300.
Results are shown in Fig.~\ref{fig:mismatches_all}.

The first row shows mismatch distributions for all simulations as a function of $n$. 
For $n=0$, we obtain distributions similar to Fig. 9 of Ref.~\cite{Boyle:2019kee} and Fig. 9 of Ref.~\cite{Scheel:2025jct}.
The simulations are highly accurate with a mean mismatch of ${<}10^{-3}$.
Comparing the generalized mismatches across $n$ reveals similar trends as the example simulations of Sec.~\ref{sec:mismatch_ex}.
The mean of the $\overline{ \mathcal{M}}_{n,0}$ distributions increases with $n$, indicating again that numerical error accumulates across the inspiral and merger. 
In contrast, the $\overline{ \mathcal{M}}_{n,n}$ distributions are consistent with each other, again suggesting that the errors in the merger are large only in relation to the earlier inspiral and not inherently.

We examine how mismatches vary across the parameter space in the three bottom rows of Fig.~\ref{fig:mismatches_all}.
In these plots, each line corresponds to one simulation. 
We separate simulations into three broad (nonexclusive) categories and use a color map to denote a pertinent parameter: eccentric simulations are colored by the eccentricity (second row), spin-aligned simulations are colored by $\chi_{\rm eff}$  (third row), and precessing simulations are colored by $\chi_p$ (bottom row).\footnote{The effective spin~\cite{Ajith:2009bn} is $\chi_{\rm eff} = (m_1 \chi_{1,||} + m_2 \chi_{2,||}) / M$ where the $||$ subscript indicates the component of each spin aligned with the Newtonian orbital angular momentum vector. The effective precessing spin~\cite{Schmidt_2015} is $\chi_{\rm p} = {\rm max}(\vec{\chi}_{1,\perp}, \vec{\chi}_{2,\perp} q (4q+3) / (4+3q) )$, where the $\perp$ subscript indicates the perpendicular component.}
We classify a simulation as eccentric if $e > 0.005$ at $t=T_{\rm ref}$, spin-aligned if $\chi_{p} \leq 10^{-4}$, and precessing if $\chi_{p}>10^{-4}$. 
Thresholds are required because spin magnitudes cannot be exactly zero in the simulation.
These classifications are not mutually exclusive such that, for example, an eccentric and precessing simulation will appear in both lists.
The broad trends identified previously persist when simulations are broken down by parameters, i.e., $\overline{ \mathcal{M}}_{n,0}$ increases as a function of $n$, while $\overline{ \mathcal{M}}_{n,n}$ has a weaker-to-no dependence. 

Turning to the parameter dependence, we find no strong trend between the value of the eccentricity and either generalized mismatch. 
This is likely because most eccentricities are still relatively low, $e\lesssim0.2$, and BBHs circularize before merger due to radiation reaction \cite{Peters:1963ux}.
This leads to a diminished imprint of eccentricity on the merger, which is precisely where we expect numerical error to be more pronounced. 

We find similar results for spin-aligned simulations and $\chi_{\rm eff}$. 
The amount of aligned spin impacts the waveform length with a larger $\chi_{\rm eff}$ leading to longer simulations~\cite{Campanelli:2006uy}.
Given that errors accumulate across the coalescence, it is therefore surprising at first glance that there is little obvious trend between accuracy and $\chi_{\rm eff}$.
The reason is that most simulations in the SXS catalog are of similar length, with a median of ${\sim} 22$ orbits~\cite{Scheel:2025jct}, imposed by the needs of the surrogate waveform model~\cite{Varma_2019}.
In other words, two simulations with different $\chi_{\rm eff}$ have different initial orbital frequencies, and thus the impact of $\chi_{\rm eff}$ on the waveform length is subdominant.
The middle panel of Fig.~\ref{fig:scatter_mismatches} shows the standard mismatch across the catalog as a function of aligned spins, again showing little correlation between accuracy and aligned spin.

There, however, is a more pronounced relationship between mismatches and the amount of precessing spin, $\chi_p$.
A larger value of $\chi_p$, and thus a larger amount of in-plane spin, correlates with larger generalized mismatches for both $\overline{ \mathcal{M}}_{n,0}$ and $\overline{ \mathcal{M}}_{n,n}$.
The same conclusion is also visible in the left panel of Fig.~\ref{fig:scatter_mismatches}, where larger in-plane spins largely correlate with greater values for $\overline{\mathcal{M}}_{0,0}$.
We suspect that this is related to the fact that \texttt{SpEC}'s grid is defined in the co-orbital frame. 
In this frame, in-plane spin components vary on the orbital timescale, whereas aligned spin components vary on the much slower radiation-reaction timescale.  So for large spin magnitudes where the near-zone geometry is highly spin-dependent, we expect it to be more difficult to resolve cases with large in-plane spins.
While the root of this inaccuracy is not the precessional motion itself, the inaccuracy is nonetheless more pronounced the higher the in-plane spin magnitude.

\section{Results: Amplitude and Phase} \label{sec:asymmetric_analysis}

In this section, we study asymmetric waveform error with the waveform and phase differences, $\Delta A^{(\ell,m)}$ and $\Delta \Phi^{(\ell,m)}$, introduced in Sec.~\ref{sec:asymmetric}. 
We first consider the results across the catalog in Sec.~\ref{sec:full_catalog_asym}, and then examine errors as a function of parameters in Sec. \ref{sec:full_catalog_asym_params}.

\subsection{Full catalog} \label{sec:full_catalog_asym}

\begin{figure*}
    \centering
    \includegraphics[width=\textwidth]{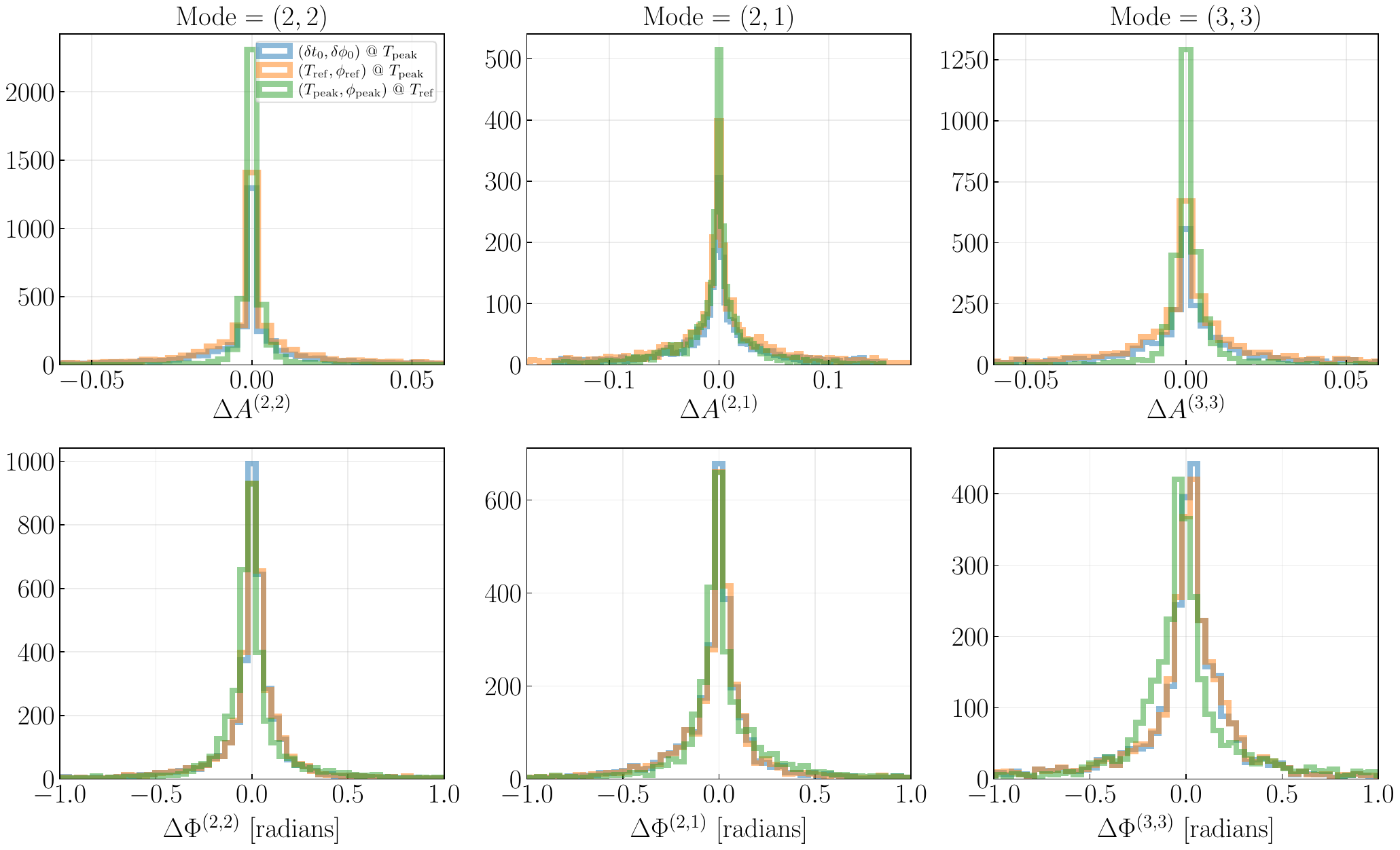}
    \caption{Histograms of the amplitude and phase differences, $\Delta A^{(\ell,m)}$ (top row) and $\Delta \Phi^{(\ell,m)}$ (bottom row), for all BBH simulations. 
    Columns correspond to the mode of interest: $(\ell,m) = (2,2)$ (left), $(2,1)$ (center), and $(3,3)$ (right).
    The legends indicate how the waveforms have been aligned and at what simulation time the quantities are evaluated.
    For example, in the green histograms waveforms have been aligned at $(T_{\rm_peak}, \phi_{\rm_peak})$ and the amplitudes and phases have been computed at time $t=T_{\rm_ref}$.
    For the dominant $(2,2)$ mode we include all simulations, whereas for the subdominant modes we apply a ``symmetry cut" of $q>1.01$, $|\chi_1-\chi_2|<0.01$, and $|\cos\theta_1-\cos\theta_2|<0.01$, where $\theta_i$ is spin tilt angle.
    Each histogram is centered about zero, indicating no noticeable systematic bias across the catalog for any modes or alignments.
    \label{fig:asymm_all}}
\end{figure*}

\begin{table*}[t]
    \centering
    \centerline{
    \begin{tabular}{|c|c|c|c|c|c|c|}\hline
   & \multicolumn{2}{|c|}{($\ell,m$)=(2,2)}& \multicolumn{2}{|c|}{($\ell,m$)=(2,1)}& \multicolumn{2}{|c|}{($\ell,m$)=(3,3)}\\
        \hline
          & 50\%& 90\%& 50\%& 90\%& 50\%& 90\%\\
          \hline \hline
     \multicolumn{7}{|c|}{$\Delta A^{(\ell,m)} $} \\
         \hline
          $(\delta t_0, \delta\phi_0)$& $(-5.5,\, 4.5)\times10^{-3}$ 
        & $(-6.5,\, 6.2)\times10^{-2}$
        & $(-2.5,\, 1.9)\times10^{-2}$
        & $(-2.4,\, 2.0)\times10^{-1}$
        & $(-10,\, 9.3) \times10^{-3}$
        & $(-11,\, 9.8)\times10^{-2}$ \\ \hline
        
         $ (T_{\rm ref}, \phi_{\rm ref})$& $(-5.5,\, 4.4)\times10^{-3}$ 
        & $(-6.5,\, 6.2)\times10^{-2} $
        & $(-2.5,\, 1.9)\times10^{-2} $
        & $(-2.3,\, 2.0)\times10^{-1} $
        & $(-9.9,\, 9.3)\times10^{-3}$
        & $(-10,\, 9.9)\times10^{-2}$ \\ \hline
        
         $ (T_{\rm peak}, \phi_{\rm peak})$& $(-9.1,\, 8.8)\times10^{-4} $
        & $(-5.2,\, 6.2)\times10^{-3} $
        & $(-13,\, 8.5)\times10^{-3} $
        & $(-8.3,\, 6.0)\times10^{-2}$ 
        & $(-1.5,\, 2.1)\times10^{-3}$ 
        & $(-7.2,\, 11)\times10^{-3}$ \\ \hline
        \hline
     \multicolumn{7}{|c|}{$\Delta \Phi^{(\ell,m)}$ [radians]}\\ 
         \hline 
          $(\delta t_0, \delta\phi_0)$& $(-3.7,\, 6.2)\times10^{-2}$ 
        & $(-4.3,\, 3.4)\times10^{-1}$ 
        & $(-8.2,\, 6.1)\times10^{-2}$ 
        & $(-5.0,\, 4.8)\times10^{-1}$ 
        & $(-7.6,\, 12)\times10^{-2}$ 
        & $(-7.7,\, 6.3)\times10^{-1}$ \\ \hline
        
        $(T_{\rm ref}, \phi_{\rm ref})$& $(-3.9,\, 6.6)\times10^{-2}$ 
        & $(-4.3,\, 3.4)\times10^{-1}$ 
        & $(-7.8,\, 6.2)\times10^{-2}$ 
        & $(-5.0,\, 4.8)\times10^{-1}$ 
        & $(-7.7,\, 13)\times10^{-2}$ 
        & $(-7.6,\, 6.9)\times10^{-1}$ \\ \hline
        
        $ (T_{\rm peak}, \phi_{\rm peak})$& $(-6.6,\, 4.9)\times10^{-2}$ 
        & $(-3.4,\, 4.2)\times10^{-1}$ 
        & $(-6.3,\, 7.9)\times10^{-2}$ 
        & $(-4.9,\, 5.0)\times10^{-1}$ 
        & $(-13,\, 7.7)\times10^{-2}$ 
        & $(-6.4,\, 7.6)\times10^{-1}$ \\ \hline
    \end{tabular}
    }
    \caption{Table of the symmetric 50\% and 90\% intervals for the distributions shown in Fig.~\ref{fig:asymm_all} for all modes and choices for alignment and evaluation time. 
}
    \label{tab:50_90_asymm_hists}
\end{table*}

We start with the 3,605 SXS simulations from Sec.~\ref{sec:mismatches_full} and plot histograms of $\Delta A^{(\ell,m)}$ and $\Delta \Phi^{(\ell,m)}$ in Fig.~\ref{fig:asymm_all}. 
Results are presented for the $(\ell,m) = (2,2)$, $(2,1)$, and $(3,3)$ waveform modes under three alignments and evaluation times.
In the first two, waveforms have been aligned by $(\delta t_0, \delta \phi_0)$ (blue) and $(T_{\rm ref}, \phi_{\rm ref})$ (orange), cf. Sec.~\ref{sec:asymmetric}, and the amplitude and phase differences are evaluated at $t=T_{\rm peak}$.
The histograms are nearly identical, as both alignments are dominated by inspiral agreement, so either effectively tests the consistency of the merger relative to the inspiral.
In what follows, we discuss one of them but conclusions apply to both.
In the third case, waveforms have been aligned at $(T_{\rm peak}, \phi_{\rm peak})$ (green), and the amplitude and phase differences are evaluated at $t=T_{\rm ref}$.
This setup assesses how accurately the early inspiral can be predicted based on the merger, the converse of the first and second cases.

Across all alignment prescriptions and modes, the amplitude and phase differences peak near zero, with average values for all distributions on the order of $10^{-3}$ and $0.04$ radians, respectively. 
This indicates that the simulation error, even mode-by-mode for the three modes considered, does not exhibit a large bias across the entire catalog and suggests that there is no systematic source of error in the numerical evolution. 
In this work, we define systematic error as errors that affect simulations in similar, i.e., systematic, ways across the parameter space.
Such errors, including catalog-wide effects from truncation error, AMR resolution effects, or gauge choices, would manifest as a consistent bias in the accuracy metrics across the catalog.
Other sources of error would be random, which would appear as random scatter in our accuracy metrics.

Turning to the error magnitude, Table~\ref{tab:50_90_asymm_hists} lists the symmetric 50\% and 90\% bounds for each distribution of Fig.~\ref{fig:asymm_all} to quantify their spread.
For the phase differences $\Delta \Phi^{(\ell, m)}$, the choice of alignment and evaluation time has minimal effect across all modes.
Roughly the same number of simulations have a positive or negative phase difference between their highest and second-highest resolutions.
For example, 90\% of the simulations have phase differences up to $4\times 10^{-1}$ radians for the $(2,2)$ mode, $5\times 10^{-1}$ radians for the $(2,1)$ mode, and $7\times 10^{-1}$ radians for the $(3,3)$ mode.
These are a fraction of an orbit, phases are therefore generally very accurate between resolutions for all modes considered.
A more detailed discussion of the relationships between mode-by-mode phase differences and their dephasing ratios is provided in Appendix \ref{apx:fractional_dephasing}. 

For the amplitude differences $\Delta A^{(\ell,m)}$, the inspiral-dominated alignments (blue and orange) are nearly identical.
On the other hand, the amplitude differences for waveforms aligned at the merger and evaluated at $t = T_{\rm ref}$ (green) are smaller by about an order of magnitude.
For the $(2,2)$ mode, 90\% of amplitude differences are below $6.5\times10^{-2}$ with inspiral-dominated alignment evaluated at merger, compared to $6.2\times10^{-3}$ when aligned at merger and evaluated at $t=T_{\rm ref}$.
This narrowing means that early inspiral behavior is more accurately predicted from the merger than vice versa. 
This is because the amplitude varies more slowly during the inspiral than during the merger, so small shifts between resolutions lead to larger amplitude differences at merger than during the inspiral.
We do not observe the same broadening when comparing $\Delta \Phi^{(2,2)}$ between $t = T_{\rm ref}$ and $t = T_{\rm peak}$.

We finally compare errors across individual modes.
For the subdominant modes $(2,1)$ and $(3,3)$, we apply a ``symmetry cut" that omits simulations with mass ratios within $1$\% of $q=1$ and spins with magnitude and direction within $1$\% of each other.
This cut targets configurations that do not excite the subdominant modes in the first place.
The dominant $(2,2)$ mode has the smallest amplitude differences.
The 90\% range increases by factors of 3.46 and 1.65 for the $(2,1)$ and $(3,3)$ modes, respectively, for inspiral-dominated alignments evaluated at $t = T_{\rm peak}$, and by factors of 12.5 and 1.59 when aligned at the merger and evaluated at $t = T_{\rm ref}$.
This broadening indicates additional error in subdominant modes, likely due to their smaller amplitudes, which are more sensitive to numerical noise.

\subsection{Parameter dependence} \label{sec:full_catalog_asym_params}

\begin{figure}
    \centering
    \includegraphics[width=\columnwidth]{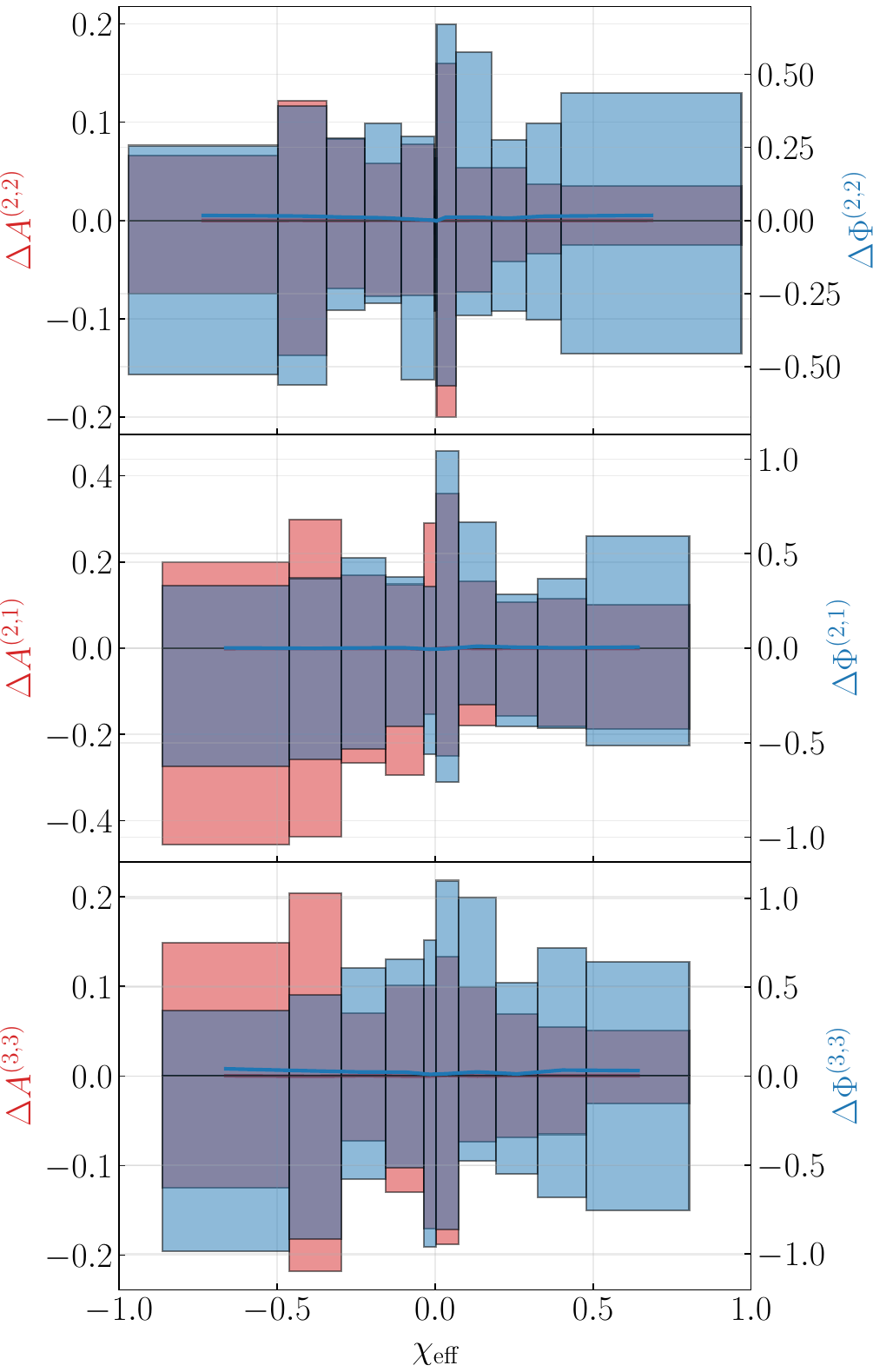}
    \caption{Medians (solid lines) and symmetric 90\% intervals (bars) on $\Delta A ^{(\ell,m)}$ (red, left vertical axis) and $\Delta \Phi ^{(\ell,m)}$ (blue, right vertical axis) for simulations with aligned spin as a function of $\chi_{\rm eff}$.
    We present results for the $(2,2)$ mode (top), $(2,1)$ mode (middle), and $(3,3)$ mode (bottom). 
    The bins are unequally spaced in $\chi_{\rm eff}$ so that each bin has approximately 300 simulations.
    The medians are close to zero and the bin sizes do not systematically vary with $\chi_{\rm eff}$, hence there is no correlation between the simulation error and aligned spin. 
    \label{fig:asymm_AS}}
\end{figure}

\begin{figure}
    \centering
    \includegraphics[width=\columnwidth]{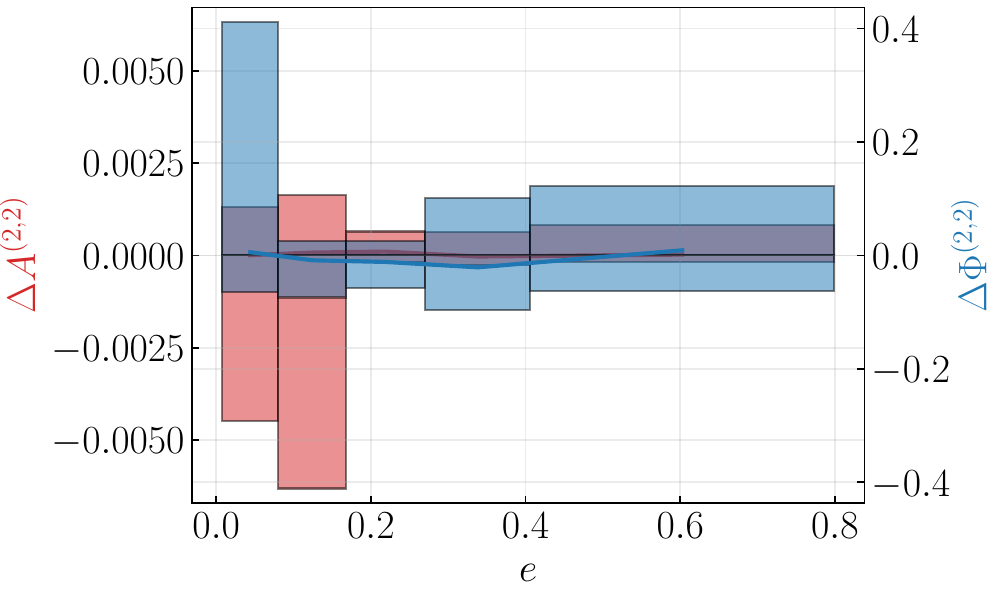}
    \caption{Similar to Fig.~\ref{fig:asymm_AS} but for eccentric simulations binned as a function of $e$ and with approximately 50 simulations per bin. 
    We restrict to the $(2,2)$ mode as the ``symmetry cuts" leave too few eccentric simulations.
    \label{fig:asymm_ecc}}
\end{figure}

\begin{figure}
    \centering
    \includegraphics[width=\columnwidth]{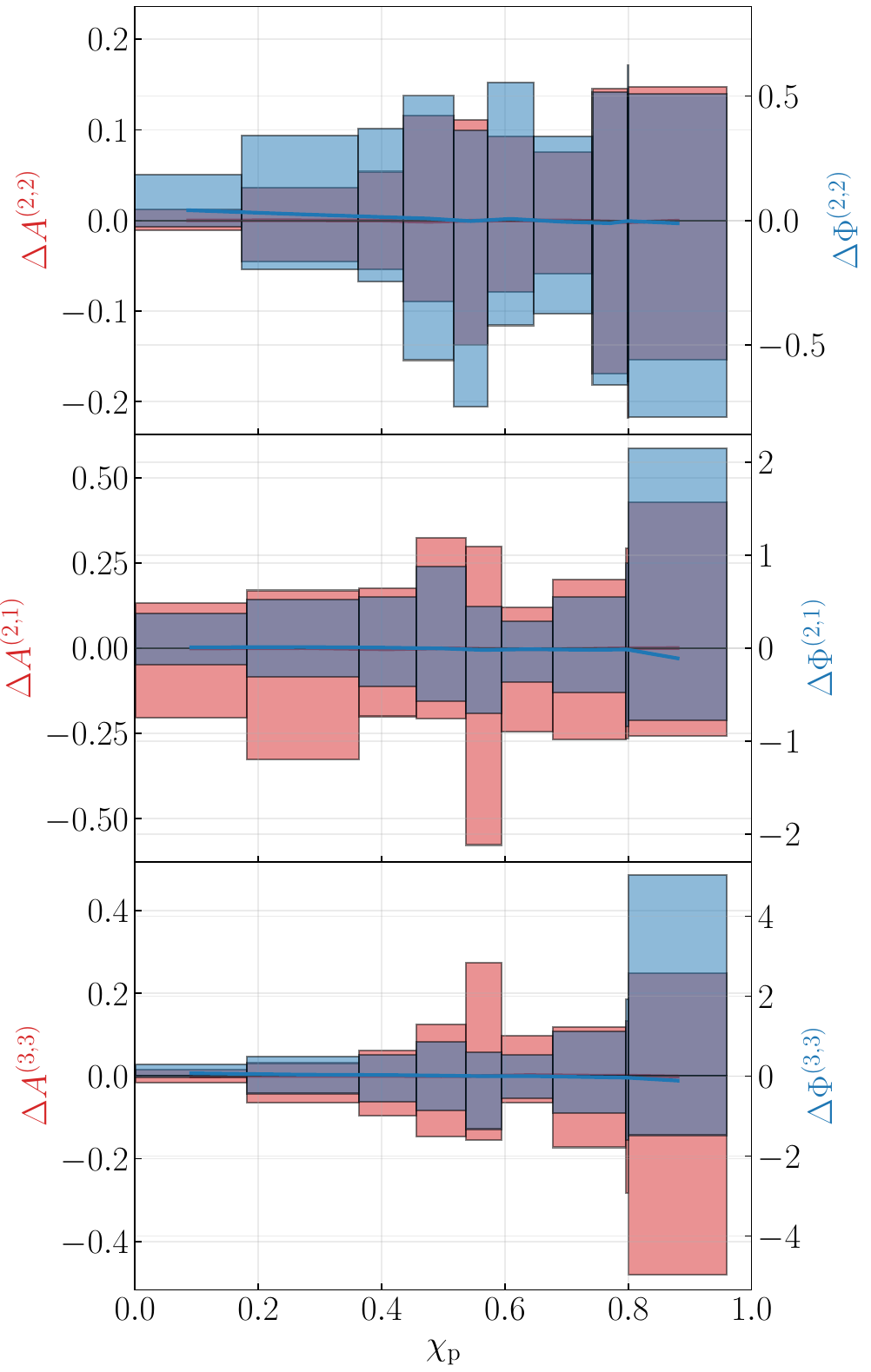}
    \caption{Similar to Fig.~\ref{fig:asymm_AS} but for precessing simulations binned as a function of $\chi_p$. 
    Although the medians are close to zero for all values of $\chi_p$, the 90\% intervals of the amplitude and phase differences broaden with increasing $\chi_p$, indicating that simulation error grows with the amount of in-plane spin.
    \label{fig:asymm_prec}}
\end{figure}

We finally examine the amplitude and phase differences as a function of the system parameters using the same nonexclusive categories as Sec.~\ref{sec:mismatches_full}. 
We show results for all modes for aligned and precessing spins in Figs.~\ref{fig:asymm_AS} and~\ref{fig:asymm_prec} respectively. 
For eccentric simulations, in Fig.~\ref{fig:asymm_ecc}, we restrict to the $(2,2)$ mode as very few eccentric simulations survive the ``symmetry cuts" outlined in Sec.~\ref{sec:full_catalog_asym}. 
Moreover, we only show the results from the $(\delta t_0, \delta \phi_0)$ waveform alignment and with quantities evaluated at $t=T_{\rm peak}$ (blue in Fig.~\ref{fig:asymm_all}) as we obtain qualitatively similar results across the different prescriptions.
Each figure shows the medians (solid lines) and symmetric 90\% intervals (bars) for the amplitude (red, left vertical axis) and phase differences (blue, right vertical axis) plotted against the parameter of interest (horizontal axis).  
The bin sizes vary such that each one contains roughly the same number of simulations: ${\sim}300$ for the quasicircular cases, and ${\sim}50$ for the eccentric case.

We begin with the aligned spin simulations shown in Fig.~\ref{fig:asymm_AS} as a function of $\chi_{\rm eff}$.
For all modes the median amplitude and phase differences remain close to zero for all values of $\chi_{\rm eff}$, reaching up to $2.3\times10^{-3}$ and $4.1\times10^{-2}$ radians respectively. 
Moreover, the 90\% intervals are roughly symmetric around zero, indicating no systematic bias in simulation error as a function of $\chi_{\rm eff}$. 
Consistent with Fig.~\ref{fig:mismatches_all}, we also find no correlation between the magnitude of the differences (height of the bins) and $\chi_{\rm eff}$.

Eccentric simulations are examined as a function of eccentricity in Fig.~\ref{fig:asymm_ecc}. 
The medians again remain close to zero, with maximum values at $1.0 \times 10^{-4}$ and $2.1 \times 10^{-2}$ radians for the amplitude and phase respectively. 
Compared to quasicircular simulations, the amplitude differences in these eccentric simulations are smaller by two orders of magnitude.
The 90\% intervals reach up to $6.3\times 10^{-3}$ and $4.1\times 10^{-1}$ radians in amplitude and phase respectively.
Although the errors also appear to increase with lower eccentricity, we find no correlation between eccentricity and simulation length that would explain this behavior.
However, given the small number of eccentric simulations available, we refrain from drawing a strong conclusion about this trend.

Finally, we turn to the precessing simulations as a function of $\chi_p$ in Fig.~\ref{fig:asymm_prec}. 
As in the aligned-spin case, we find no evidence of bias.
Medians errors are small, with maxima of $2.0\times10^{-3}$ and $4.1\times10^{-2}$ radians for the $(2,2)$ mode,  $3.3\times10^{-3}$ and $1.1\times10^{-1}$ radians for $(2,1)$, and $2.3\times10^{-3}$ and $1.2\times10^{-1}$ radians for $(3,3)$.
Bars are symmetric, again with no evidence for a bias.
However, and consistent with Fig.~\ref{fig:mismatches_all} and the discussion of Sec.~\ref{sec:mismatches_full}, for both the amplitude and the phase differences, there is a clear correlation between $\chi_p$ and the simulation error.
This is more pronounced for the $(2,2)$ and $(3,3)$ modes, but also clear in the $(2,1)$ mode phase.
The widths of the 90\% intervals for amplitude and phase are broadened by factors up to $15.1$ and $6.2$ for the $(2,2)$ mode,  $2.6$ and $5.3$ for $(2,1)$, and $23.1$ and $19.8$ for $(3,3)$, respectively, as $\chi_p$ increases.
Our interpretation is the same as in Sec.~\ref{sec:mismatches_full}:
increased errors reflect the increased morphological complexity of precessing waveforms.

\section{Conclusions} \label{sec:discussion}

In this work, we proposed three accuracy metrics of NR waveforms and applied them to the SXS catalog: the generalized frequency-weighted mismatch $\overline{\mathcal M}_{n,n}$, the relative amplitude difference $\Delta A^{(\ell,m)}$, and the phase difference $\Delta \Phi^{(\ell,m)}$. 
The generalized mismatch allows us to target different portions of the waveform and shows that although error accumulates during the system evolution, the merger and ringdown stages are not inherently less accurate. 
The amplitude and phase differences allow us to probe for systematic simulation errors, i.e., errors that manifest qualitatively similarly across the parameter space, for which we find no evidence. 
This extensive accuracy assessment of the SXS catalog represents the most thorough evaluation of the largest NR waveform catalog available to date.
While here we focus on the SXS catalog, the metrics and analyzes presented are equally applicable to other catalogs ~\cite{Scheel:2025jct, ferguson2023secondmayacatalogbinary, Healy_2020, Rashti_2025, hamilton2023catalogueprecessingblackholebinarynumericalrelativity, Huerta_2019, Hinder:2013oqa, Aylott:2009tn}.

By conducting a parameter space study, we found that
across all three metrics, errors increase with in-plane spin, reflecting the higher morphological complexity of precessing systems and their waveforms. 
The \texttt{SpEC} grid is defined in the co-orbital frame, which causes the geometry of the near-zone of the simulation to be highly spin-dependent. 
As a result, it is more difficult to resolve systems with greater in-plane spin, resulting in larger simulation errors. 
By identifying how greater in-plane spin leads to larger simulation errors due to the co-orbital grid frame, we hope to motivate the development of methods to mitigate this effect in future versions of \texttt{SpEC}.

The three accuracy metrics introduced here not only assess NR waveform accuracy within the SXS catalog but can also quantify errors in downstream targeted applications. 
For example, QNM studies or IMR consistency tests should consider different generalized mismatch alignments to assess the relevant errors.
QNM studies focus on the rigndown signal in isolation. 
The upweighted alignments, i.e., $(\delta t_n, \delta \phi_n)$, are therefore more relevant as they focus on the ringdown-dominated, higher-frequency regime and avoid effects from inspiral-driven effects at lower-frequency regimes of the waveform.
Alternatively, IMR consistency tests are related to mismatches with one-time alignment, i.e. $(\delta t_0, \delta \phi_0)$, that preserve the connection between early and late times in the waveform when comparing inferred parameters. 
Another potential application for these accuracy metrics is in evaluating surrogate errors. 
The asymmetric accuracy metrics enable mode-by-mode comparisons between the surrogate and its NR training waveforms, assisting in the identification of less accurate regimes.
More broadly, though, additional metrics and systematic explorations tailored to specific \texttt{SpEC} data products and their applications, such as the remnant properties or QNM fits, are required given the central role of NR in GW science. 


\acknowledgements

We thank Saul Teukolsky for discussions about the mismatch and simulation errors.
We also thank Kyle Nelli and the SXS group for advice on using \texttt{SpEC} and the SXS catalog. 
The computations presented here were conducted at the Resnick High Performance Computing Center, a facility supported by Resnick Sustainability Institute at the California Institute of Technology. 
This work was supported by the Sherman Fairchild Foundation, and by NSF Grants No. PHY-2309211, No. PHY-2309231, and No. OAC2209656 at Caltech. K.M. is supported by NASA through the NASA Hubble Fellowship Grant No. HST-HF2-51562.001-A awarded by the Space Telescope Science Institute, which is operated by the Association of Universities for Research in Astronomy, Incorporated, under NASA contract NAS5-26555.
This material is based upon work supported by the National Science Foundation under Grants No. PHY-2407742, No. PHY- 2207342, and No. OAC-2209655, and  by the Sherman Fairchild Foundation at Cornell.

The analysis presented in this work made use of a number of open-source software packages. 
Data processing and numerical computations were performed using \texttt{NumPy}~\cite{Harris_2020}, \texttt{SciPy}~\cite{2020SciPy-NMeth}, and \texttt{matplotlib}~\cite{Hunter:2007}. 
In particular, we acknowledge the use of the \texttt{sxs}~\cite{boyle2025sxs} and \texttt{scri}~\cite{boyle2025scri} packages, which were developed to analyze numerical relativity simulation data and waveforms. 
The \texttt{SpEC} code~\cite{SpEC} was utilized for the simulations in the SXS Catalog, which is the proprietary SXS numerical relativity code. 
The authors gratefully acknowledge the developers and maintainers of these projects for their positive impact on the scientific community.

\appendix

\section{$\hat{L}^2$ minimization over $(\delta t, \delta \phi)$}
\label{apx:L2_minimization}

As discussed in Sec.~\ref{sec:mismatches}, before computing the mismatch between two waveforms, we must align them in time and phase. 
However, directly minimizing the \textit{mismatch} over time and phase is insensitive to an overall scaling difference.
To circumvent this, we follow Ref.~\cite{Scheel:2025jct} and minimize the normalized $L^2$ norm, $\hat{L}^2$, between the two waveforms instead.

The $L^2$ norm between two waveforms is
\begin{alignat}{2} 
    L^2& =( h^{\rm I} - h^{\rm II}(\delta t, \delta \phi) | h^{\rm I} - h^{\rm II}(\delta t, \delta \phi)) \label{eq:L2_norm2} \,,
\end{alignat}
where waveform $h^{\rm II}$ has been translated by some time shift, $\delta t$, and azimuthally rotated by some angle, $\delta \phi$.
We formulate the inner product as the integral
\begin{equation}
    ( a|b ) = 4 {\rm Re}\int\int_{S^2} a(t)b^*(t)d\Omega dt\,,
\end{equation}
where we integrate over time and the two-sphere described by $(\theta, \phi)$. 
Inserting Eq.~\eqref{eqn:full_h} into Eq.~\eqref{eq:L2_norm2} yields
\begin{alignat}{2} \label{eq:intermediate_int}
    L^2 &= 4 {\rm Re} \int^{t_2}_{t_1} \int_{S^2}  \left|\sum_{\ell, m}h^{(\ell,m)}_{\rm I}(t)_{-2}Y_{(\ell, m)}(\theta, \phi) \nonumber \right.\\
    &\left.- h^{(\ell,m)}_{\rm II}(t+\delta t)_{-2}Y_{(\ell, m)}(\theta, \phi+\delta \phi)\right|^2 d\Omega dt \,.
\end{alignat}
We integrate from $t=t_1$ to $t=t_2$, such that $t_1$ is the later $T_{\rm ref}$ of the two waveforms and $t_2$ is late enough that ringdown has concluded (in practice, the end of the data).
The phase shift, $\delta \phi$, separates from each spin-weighted spherical harmonic, allowing Eq.~\eqref{eq:intermediate_int} to be separated into two integrals
\begin{alignat}{2} \label{eq:intermediate_int2}
    L^2 &= 4 {\rm Re} \sum_{\ell, m} \int_{S^2} \left|_{-2}Y_{(\ell, m)}(\theta, \phi)\right|^2 d\Omega\, \nonumber \\ &\times \, \int^{t_2}_{t_1} \left|h^{(\ell,m)}_{\rm I}(t) - h^{(\ell,m)}_{\rm II}(t+\delta t)e^{im\delta \phi}\right|^2 dt \,.
\end{alignat}
Spin-weighted spherical harmonics are orthogonal, so we rewrite the angular portion of Eq.~\eqref{eq:intermediate_int2} as
\begin{align}
    \int_{S^2} &\left|_{-2}Y_{(\ell, m)}(\theta, \phi)\right|^2 d\Omega \nonumber\\
    &= \int_{S^2} \,_{-2}Y_{(\ell, m)}(\theta, \phi) _{-2}Y^*_{(\ell', m')}(\theta, \phi) d\Omega \nonumber \\ 
    &= \delta_{\ell, \ell'} \delta_{m, m'}\,.
\end{align}
This forces $\ell' = \ell$ and $m'=m$, so Eq.~\eqref{eq:intermediate_int2} reduces to
\begin{alignat}{2} \label{eq:L2_final}
    L^2 &= 4 {\rm Re} \sum_{\ell, m} \int^{t_2}_{t_1} \left|h^{(\ell,m)}_{\rm I}(t) - h^{(\ell,m)}_{\rm II}(t+\delta t)e^{im\delta \phi}\right|^2 dt \,,
\end{alignat}
that is independent of $\theta$ and $\phi$. 
Finally, the normalized $L^2$ norm is
\begin{alignat}{2}
    \hat{L}^2
    &= \frac{( h^{\rm I} - h^{\rm II}(\delta t, \delta \phi) | h^{\rm I} - h^{\rm II}(\delta t, \delta \phi))}{( h^{\rm I} | h^{\rm I} )} \nonumber \\
    &= \frac{\sum_{\ell, m}\int^{t_2}_{t_1}\left|h^{(\ell,m)}_{\rm I}(t) - h^{(\ell,m)}_{\rm II}(t+\delta t)e^{im\delta \phi}\right|^2 dt}{\sum_{\ell, m} \int^{t_2}_{t_1}\left|h^{(\ell,m)}_{\rm I}(t)\right|^2 dt}\,.
    \label{eq:L2_hat_final}
\end{alignat}
Equation~\eqref{eq:L2_hat_final} is what we minimize to obtain the time and phase shifts, $\delta t$ and $\delta \phi$.

\section{Ratio of dephasings}
\label{apx:fractional_dephasing}

\begin{figure}
    \centering
    \includegraphics[width=\columnwidth]{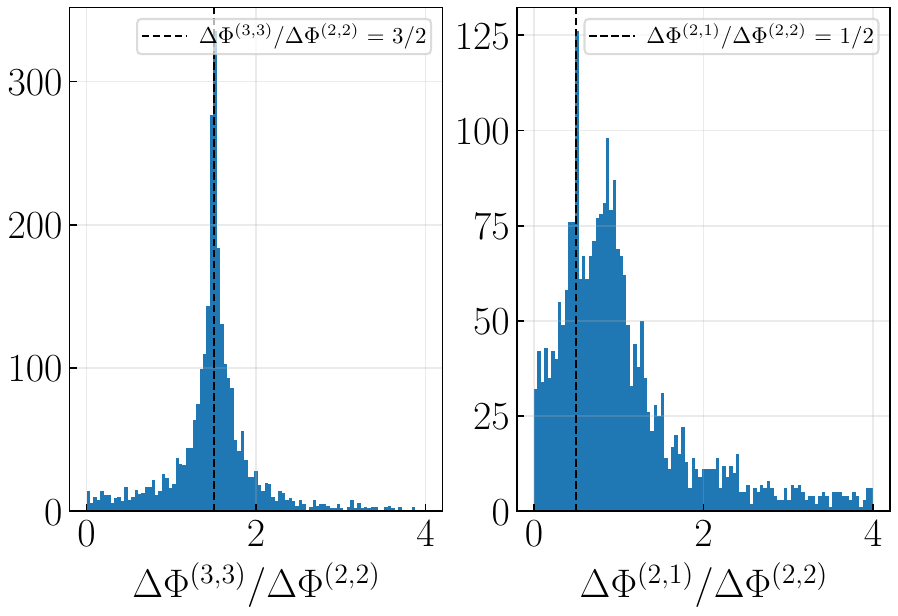}
    \caption{Histograms of the ratios between the phase differences of three modes of interest computed in Sec. \ref{sec:asymmetric_analysis}.
    For the subset of simulations that satisfy the cuts defined in Sec. \ref{sec:asymmetric_analysis}, we show the ratios of the dephasing between the $(2,2)$ and $(3,3)$ modes (left), and the $(2,2)$ and $(2,1)$ modes (right). 
    These phase differences are computed using the $(\delta t_0, \delta \phi_0)$ alignment procedure and evaluated at $t=T_{\rm peak}$.
    The expected first-order dephasing ratios are shown in vertical black dashed lines. 
    The histograms for each dephasing ratio are centered around and peak at the expected first-order dephasing ratio values, which verifies this first-order relationship. 
    Any deviation from the expected first-order dephasing ratios may be attributed to spin and eccentricity corrections to the mode phases at higher PN order, or the difference in junk radiation between simulation resolutions.
    \label{fig:frac_dephasing}}
\end{figure}

To leading order in post-Newtonian (PN) theory, the phase for the $(\ell, m)$ mode as a function of time is related to the orbital phase $\phi_{\rm orb}(t)$ via~\cite{Kidder:2007rt}
\begin{equation}
    \Phi^{(\ell, m)} (t) \simeq m \phi_{\rm orb} (t)\,.
\end{equation}
Two different resolutions of the same NR simulation have a slightly different orbital phase at late times (such as at merger) due to numerical errors, which suggests a dephasing of the modes given by
\begin{equation}
    \Delta \Phi^{(\ell,m)}(t) \simeq m (\phi_{\rm orb}^I (t) - \phi_{\rm orb}^{II} (t))\,.
\end{equation}
The superscripts $I$ and $II$ correspond to the first and second highest resolutions, respectively.
Therefore, we expect that, to first order, the dephasing between modes is related via a factor of $m$. 

We can use the phase differences computed in Sec. \ref{sec:asymmetric_analysis} to verify this first-order dephasing relationship. 
If this first-order dephasing is true, then the ratios between the $(\ell, m)$ and $(\ell', m')$ phase differences should satisfy: 
\begin{equation} \label{eq:frac_dephase}
    \frac{\Delta \Phi^{(\ell,m)}(t) }{\Delta \Phi^{(\ell',m')}(t)} \simeq \frac{m}{m'}\,.
\end{equation}

For each simulation in the SXS catalog that satisfies our cuts, we plot the ratios between the phase differences in Fig. \ref{fig:frac_dephasing}.
We also plot the expected first-order dephasing ratio for each pair of modes as computed in Eq.~\eqref{eq:frac_dephase} (black dashed lines).
Because each histogram peaks at its expected dephasing value, we confirm that our results are consistent with first-order expectations. 

The tails on the histograms may be attributed to a few effects. 
One reason is that nonzero spins and eccentricity enter PN expressions for the phase differently across modes, causing deviations in the mode dephasings from the expected first-order ratio~\cite{Henry:2022dzx}. 
Another reason is that junk radiation at the beginning of the simulation is poorly resolved and can affect different resolutions of the same simulation in different ways, introducing a small effectively random initial perturbation into each resolution.
Although there are controls in \texttt{SpEC} to mitigate this effect~\cite{Scheel:2025jct}, slight differences in the same mode at different resolutions can result in dephasings that deviate from the expected first-order ratio.

\bibliography{Refs}

@article{Boyle:2019kee,
    author = "Boyle, Michael and others",
    title = "{The SXS Collaboration catalog of binary black hole simulations}",
    eprint = "1904.04831",
    archivePrefix = "arXiv",
    primaryClass = "gr-qc",
    doi = "10.1088/1361-6382/ab34e2",
    journal = "Class. Quant. Grav.",
    volume = "36",
    number = "19",
    pages = "195006",
    year = "2019"
}

@misc{SXSCatalog,
  howpublished = "\url{http://www.black-holes.org/waveforms}"
}

@misc{SpEC,
  howpublished = "\url{https://www.black-holes.org/for-researchers/spec}"
}

@article{Garcia-Quiros:2020qpx,
    author = "Garc{\'\i}a-Quir{\'o}s, Cecilio and Colleoni, Marta and Husa, Sascha and Estell{\'e}s, H{\'e}ctor and Pratten, Geraint and Ramos-Buades, Antoni and Mateu-Lucena, Maite and Jaume, Rafel",
    title = "{Multimode frequency-domain model for the gravitational wave signal from nonprecessing black-hole binaries}",
    eprint = "2001.10914",
    archivePrefix = "arXiv",
    primaryClass = "gr-qc",
    doi = "10.1103/PhysRevD.102.064002",
    journal = "Phys. Rev. D",
    volume = "102",
    number = "6",
    pages = "064002",
    year = "2020"
}

@article{Estelles:2025zah,
    author = "Estell{\'e}s, H{\'e}ctor and Buonanno, Alessandra and Enficiaud, Raffi and Foo, Cheng and Pompili, Lorenzo",
    title = "{Adding equatorial-asymmetric effects for spin-precessing binaries into the SEOBNRv5PHM waveform model}",
    eprint = "2506.19911",
    archivePrefix = "arXiv",
    primaryClass = "gr-qc",
    month = "6",
    year = "2025"
}

@article{Ghosh:2023mhc,
    author = "Ghosh, Shrobana and Kolitsidou, Panagiota and Hannam, Mark",
    title = "{First frequency-domain phenomenological model of the multipole asymmetry in gravitational-wave signals from binary-black-hole coalescence}",
    eprint = "2310.16980",
    archivePrefix = "arXiv",
    primaryClass = "gr-qc",
    doi = "10.1103/PhysRevD.109.024061",
    journal = "Phys. Rev. D",
    volume = "109",
    number = "2",
    pages = "024061",
    year = "2024"
}

@article{Thompson:2023ase,
    author = "Thompson, Jonathan E. and Hamilton, Eleanor and London, Lionel and Ghosh, Shrobana and Kolitsidou, Panagiota and Hoy, Charlie and Hannam, Mark",
    title = "{PhenomXO4a: a phenomenological gravitational-wave model for precessing black-hole binaries with higher multipoles and asymmetries}",
    eprint = "2312.10025",
    archivePrefix = "arXiv",
    primaryClass = "gr-qc",
    reportNumber = "LIGO-P2300437",
    doi = "10.1103/PhysRevD.109.063012",
    journal = "Phys. Rev. D",
    volume = "109",
    number = "6",
    pages = "063012",
    year = "2024"
}

@article{Bhat:2025lri,
    author = "Bhat, Sajad A. and Tiwari, Avinash and Shaikh, Md Arif and Kapadia, Shasvath J.",
    title = "{EECT: an Eccentricity Evolution Consistency Test to distinguish eccentric gravitational-wave signals from eccentricity mimickers}",
    eprint = "2508.14850",
    archivePrefix = "arXiv",
    primaryClass = "gr-qc",
    month = "8",
    year = "2025"
}

@article{Nagar:2020pcj,
    author = "Nagar, Alessandro and Riemenschneider, Gunnar and Pratten, Geraint and Rettegno, Piero and Messina, Francesco",
    title = "{Multipolar effective one body waveform model for spin-aligned black hole binaries}",
    eprint = "2001.09082",
    archivePrefix = "arXiv",
    primaryClass = "gr-qc",
    doi = "10.1103/PhysRevD.102.024077",
    journal = "Phys. Rev. D",
    volume = "102",
    number = "2",
    pages = "024077",
    year = "2020"
}

@article{Chiaramello:2020ehz,
    author = "Chiaramello, Danilo and Nagar, Alessandro",
    title = "{Faithful analytical effective-one-body waveform model for spin-aligned, moderately eccentric, coalescing black hole binaries}",
    eprint = "2001.11736",
    archivePrefix = "arXiv",
    primaryClass = "gr-qc",
    doi = "10.1103/PhysRevD.101.101501",
    journal = "Phys. Rev. D",
    volume = "101",
    number = "10",
    pages = "101501",
    year = "2020"
}

@article{Hamilton:2021pkf,
    author = "Hamilton, Eleanor and London, Lionel and Thompson, Jonathan E. and Fauchon-Jones, Edward and Hannam, Mark and Kalaghatgi, Chinmay and Khan, Sebastian and Pannarale, Francesco and Vano-Vinuales, Alex",
    title = "{Model of gravitational waves from precessing black-hole binaries through merger and ringdown}",
    eprint = "2107.08876",
    archivePrefix = "arXiv",
    primaryClass = "gr-qc",
    doi = "10.1103/PhysRevD.104.124027",
    journal = "Phys. Rev. D",
    volume = "104",
    number = "12",
    pages = "124027",
    year = "2021"
}

@article{Higginbotham:2019wbx,
    author = "Higginbotham, Kenny and Khamesra, Bhavesh and McInerney, Jame P. and Jani, Karan and Shoemaker, Deirdre M. and Laguna, Pablo",
    title = "{Coping with spurious radiation in binary black hole simulations}",
    eprint = "1907.00027",
    archivePrefix = "arXiv",
    primaryClass = "gr-qc",
    doi = "10.1103/PhysRevD.100.081501",
    journal = "Phys. Rev. D",
    volume = "100",
    number = "8",
    pages = "081501",
    year = "2019"
}

@article{Mezzasoma:2025moh,
    author = "Mezzasoma, Simone and Haster, Carl-Johan and Owen, Caroline B. and Cornish, Neil J. and Yunes, Nicol{\'a}s",
    title = "{Uncertainty-aware waveform modeling for high signal-to-noise ratio gravitational-wave inference}",
    eprint = "2503.23304",
    archivePrefix = "arXiv",
    primaryClass = "gr-qc",
    doi = "10.1103/hfq5-gryy",
    journal = "Phys. Rev. D",
    volume = "112",
    number = "4",
    pages = "044057",
    year = "2025"
}

@article{Owen:2023mid,
    author = "Owen, Caroline B. and Haster, Carl-Johan and Perkins, Scott and Cornish, Neil J. and Yunes, Nicol{\'a}s",
    title = "{Waveform accuracy and systematic uncertainties in current gravitational wave observations}",
    eprint = "2301.11941",
    archivePrefix = "arXiv",
    primaryClass = "gr-qc",
    doi = "10.1103/PhysRevD.108.044018",
    journal = "Phys. Rev. D",
    volume = "108",
    number = "4",
    pages = "044018",
    year = "2023"
}

@article{Tiwari:2025fua,
    author = "Tiwari, Avinash and Bhat, Sajad A. and Shaikh, MD Arif and Kapaida, Shashvath J.",
    title = "{Testing the nature of GW200105 by probing the frequency evolution of eccentricity}",
    eprint = "2509.26152",
    archivePrefix = "arXiv",
    primaryClass = "astro-ph.HE",
    month = "9",
    year = "2025"
}

@Article{Hunter:2007,
  Author    = {Hunter, J. D.},
  Title     = {Matplotlib: A 2D graphics environment},
  Journal   = {Computing in Science \& Engineering},
  Volume    = {9},
  Number    = {3},
  Pages     = {90--95},
  publisher = {IEEE COMPUTER SOC},
  doi       = {10.1109/MCSE.2007.55},
  year      = 2007
}

@ARTICLE{2020SciPy-NMeth,
  author  = {Virtanen, Pauli and Gommers, Ralf and Oliphant, Travis E. and
            Haberland, Matt and Reddy, Tyler and Cournapeau, David and
            Burovski, Evgeni and Peterson, Pearu and Weckesser, Warren and
            Bright, Jonathan and {van der Walt}, St{\'e}fan J. and
            Brett, Matthew and Wilson, Joshua and Millman, K. Jarrod and
            Mayorov, Nikolay and Nelson, Andrew R. J. and Jones, Eric and
            Kern, Robert and Larson, Eric and Carey, C J and
            Polat, {\.I}lhan and Feng, Yu and Moore, Eric W. and
            {VanderPlas}, Jake and Laxalde, Denis and Perktold, Josef and
            Cimrman, Robert and Henriksen, Ian and Quintero, E. A. and
            Harris, Charles R. and Archibald, Anne M. and
            Ribeiro, Ant{\^o}nio H. and Pedregosa, Fabian and
            {van Mulbregt}, Paul and {SciPy 1.0 Contributors}},
  title   = {{{SciPy} 1.0: Fundamental Algorithms for Scientific
            Computing in Python}},
  journal = {Nature Methods},
  year    = {2020},
  volume  = {17},
  pages   = {261--272},
  adsurl  = {https://rdcu.be/b08Wh},
  doi     = {10.1038/s41592-019-0686-2},
}

@article{Harris_2020,
   title={Array programming with NumPy},
   volume={585},
   ISSN={1476-4687},
   url={http://dx.doi.org/10.1038/s41586-020-2649-2},
   DOI={10.1038/s41586-020-2649-2},
   number={7825},
   journal={Nature},
   publisher={Springer Science and Business Media LLC},
   author={Harris, Charles R. and Millman, K. Jarrod and van der Walt, Stéfan J. and Gommers, Ralf and Virtanen, Pauli and Cournapeau, David and Wieser, Eric and Taylor, Julian and Berg, Sebastian and Smith, Nathaniel J. and Kern, Robert and Picus, Matti and Hoyer, Stephan and van Kerkwijk, Marten H. and Brett, Matthew and Haldane, Allan and del Río, Jaime Fernández and Wiebe, Mark and Peterson, Pearu and Gérard-Marchant, Pierre and Sheppard, Kevin and Reddy, Tyler and Weckesser, Warren and Abbasi, Hameer and Gohlke, Christoph and Oliphant, Travis E.},
   year={2020},
   month=sep, pages={357–362} }

@article{Boyle_2007,
   title={High-accuracy comparison of numerical relativity simulations with post-Newtonian expansions},
   volume={76},
   ISSN={1550-2368},
   url={http://dx.doi.org/10.1103/PhysRevD.76.124038},
   DOI={10.1103/physrevd.76.124038},
   number={12},
   journal={Physical Review D},
   publisher={American Physical Society (APS)},
   author={Boyle, Michael and Brown, Duncan A. and Kidder, Lawrence E. and Mroué, Abdul H. and Pfeiffer, Harald P. and Scheel, Mark A. and Cook, Gregory B. and Teukolsky, Saul A.},
   year={2007},
   month=dec }

@article{Hinder_2010,
   title={Comparisons of eccentric binary black hole simulations with post-Newtonian models},
   volume={82},
   ISSN={1550-2368},
   url={http://dx.doi.org/10.1103/PhysRevD.82.024033},
   DOI={10.1103/physrevd.82.024033},
   number={2},
   journal={Physical Review D},
   publisher={American Physical Society (APS)},
   author={Hinder, Ian and Herrmann, Frank and Laguna, Pablo and Shoemaker, Deirdre},
   year={2010},
   month=jul }

@article{Nobili:2025ydt,
    author = "Nobili, Francesco and Bhagwat, Swetha and Pacilio, Costantino and Gerosa, Davide",
    title = "{Ringdown mode amplitudes of precessing binary black holes}",
    eprint = "2504.17021",
    archivePrefix = "arXiv",
    primaryClass = "gr-qc",
    month = "4",
    year = "2025"
}

@article{Chu:2015kft,
    author = "Chu, Tony and Fong, Heather and Kumar, Prayush and Pfeiffer, Harald P. and Boyle, Michael and Hemberger, Daniel A. and Kidder, Lawrence E. and Scheel, Mark A. and Szilagyi, Bela",
    title = "{On the accuracy and precision of numerical waveforms: Effect of waveform extraction methodology}",
    eprint = "1512.06800",
    archivePrefix = "arXiv",
    primaryClass = "gr-qc",
    doi = "10.1088/0264-9381/33/16/165001",
    journal = "Class. Quant. Grav.",
    volume = "33",
    number = "16",
    pages = "165001",
    year = "2016"
}

@article{LIGOScientific:2019hgc,
    author = "Abbott, Benjamin P and others",
    collaboration = "LIGO Scientific, Virgo",
    title = "{A guide to LIGO\textendash{}Virgo detector noise and extraction of transient gravitational-wave signals}",
    eprint = "1908.11170",
    archivePrefix = "arXiv",
    primaryClass = "gr-qc",
    doi = "10.1088/1361-6382/ab685e",
    journal = "Class. Quant. Grav.",
    volume = "37",
    number = "5",
    pages = "055002",
    year = "2020"
}

@article{Chatziioannou:2017tdw,
    author = "Chatziioannou, Katerina and Klein, Antoine and Yunes, Nicol\'as and Cornish, Neil",
    title = "{Constructing Gravitational Waves from Generic Spin-Precessing Compact Binary Inspirals}",
    eprint = "1703.03967",
    archivePrefix = "arXiv",
    primaryClass = "gr-qc",
    doi = "10.1103/PhysRevD.95.104004",
    journal = "Phys. Rev. D",
    volume = "95",
    number = "10",
    pages = "104004",
    year = "2017"
}

@article{Lindblom:2008cm,
    author = "Lindblom, Lee and Owen, Benjamin J. and Brown, Duncan A.",
    title = "{Model Waveform Accuracy Standards for Gravitational Wave Data Analysis}",
    eprint = "0809.3844",
    archivePrefix = "arXiv",
    primaryClass = "gr-qc",
    doi = "10.1103/PhysRevD.78.124020",
    journal = "Phys. Rev. D",
    volume = "78",
    pages = "124020",
    year = "2008"
}

@article{Udall:2024ovp,
    author = "Udall, Rhiannon and Hourihane, Sophie and Miller, Simona and Davis, Derek and Chatziioannou, Katerina and Isi, Max and Deshong, Howard",
    title = "{Antialigned spin of GW191109: Glitch mitigation and its implications}",
    eprint = "2409.03912",
    archivePrefix = "arXiv",
    primaryClass = "gr-qc",
    doi = "10.1103/PhysRevD.111.024046",
    journal = "Phys. Rev. D",
    volume = "111",
    number = "2",
    pages = "024046",
    year = "2025"
}

@article{Ghosh:2017gfp,
    author = "Ghosh, Abhirup and Johnson-Mcdaniel, Nathan K. and Ghosh, Archisman and Mishra, Chandra Kant and Ajith, Parameswaran and Del Pozzo, Walter and Berry, Christopher P. L. and Nielsen, Alex B. and London, Lionel",
    title = "{Testing general relativity using gravitational wave signals from the inspiral, merger and ringdown of binary black holes}",
    eprint = "1704.06784",
    archivePrefix = "arXiv",
    primaryClass = "gr-qc",
    reportNumber = "LIGO-P1700006, ICTS-2017-3",
    doi = "10.1088/1361-6382/aa972e",
    journal = "Class. Quant. Grav.",
    volume = "35",
    number = "1",
    pages = "014002",
    year = "2018"
}

@article{Isi:2020tac,
    author = "Isi, Maximiliano and Farr, Will M. and Giesler, Matthew and Scheel, Mark A. and Teukolsky, Saul A.",
    title = "{Testing the Black-Hole Area Law with GW150914}",
    eprint = "2012.04486",
    archivePrefix = "arXiv",
    primaryClass = "gr-qc",
    reportNumber = "LIGO-P2000507",
    doi = "10.1103/PhysRevLett.127.011103",
    journal = "Phys. Rev. Lett.",
    volume = "127",
    number = "1",
    pages = "011103",
    year = "2021"
}

@article{Pacilio:2024tdl,
    author = "Pacilio, Costantino and Bhagwat, Swetha and Nobili, Francesco and Gerosa, Davide",
    title = "{Flexible mapping of ringdown amplitudes for nonprecessing binary black holes}",
    eprint = "2408.05276",
    archivePrefix = "arXiv",
    primaryClass = "gr-qc",
    doi = "10.1103/PhysRevD.110.103037",
    journal = "Phys. Rev. D",
    volume = "110",
    number = "10",
    pages = "103037",
    year = "2024"
}

@article{Ghosh:2016qgn,
    author = "Ghosh, Abhirup and others",
    title = "{Testing general relativity using golden black-hole binaries}",
    eprint = "1602.02453",
    archivePrefix = "arXiv",
    primaryClass = "gr-qc",
    reportNumber = "LIGO-P1500185-V10, ICTS-2016-1, LIGO-P1500185-V11",
    doi = "10.1103/PhysRevD.94.021101",
    journal = "Phys. Rev. D",
    volume = "94",
    number = "2",
    pages = "021101",
    year = "2016"
}

@article{MaganaZertuche:2024ajz,
    author = "Maga{\~n}a Zertuche, Lorena and others",
    title = "{High-precision ringdown surrogate model for nonprecessing binary black holes}",
    eprint = "2408.05300",
    archivePrefix = "arXiv",
    primaryClass = "gr-qc",
    doi = "10.1103/q7sy-g3kl",
    journal = "Phys. Rev. D",
    volume = "112",
    number = "2",
    pages = "024077",
    year = "2025"
}

@article{LIGOScientific:2019fpa,
    author = "Abbott, B. P. and others",
    collaboration = "LIGO Scientific, Virgo",
    title = "{Tests of General Relativity with the Binary Black Hole Signals from the LIGO-Virgo Catalog GWTC-1}",
    eprint = "1903.04467",
    archivePrefix = "arXiv",
    primaryClass = "gr-qc",
    reportNumber = "LIGO-P1800316",
    doi = "10.1103/PhysRevD.100.104036",
    journal = "Phys. Rev. D",
    volume = "100",
    number = "10",
    pages = "104036",
    year = "2019"
}

@article{LIGOScientific:2016lio,
    author = "Abbott, B. P. and others",
    collaboration = "LIGO Scientific, Virgo",
    title = "{Tests of general relativity with GW150914}",
    eprint = "1602.03841",
    archivePrefix = "arXiv",
    primaryClass = "gr-qc",
    reportNumber = "LIGO-P1500213",
    doi = "10.1103/PhysRevLett.116.221101",
    journal = "Phys. Rev. Lett.",
    volume = "116",
    number = "22",
    pages = "221101",
    year = "2016",
    note = "[Erratum: Phys.Rev.Lett. 121, 129902 (2018)]"
}

@article{LIGOScientific:2020tif,
    author = "Abbott, R. and others",
    collaboration = "LIGO Scientific, Virgo",
    title = "{Tests of general relativity with binary black holes from the second LIGO-Virgo gravitational-wave transient catalog}",
    eprint = "2010.14529",
    archivePrefix = "arXiv",
    primaryClass = "gr-qc",
    reportNumber = "LIGO-P2000091",
    doi = "10.1103/PhysRevD.103.122002",
    journal = "Phys. Rev. D",
    volume = "103",
    number = "12",
    pages = "122002",
    year = "2021"
}

@article{LIGOScientific:2018dkp,
    author = "Abbott, B. P. and others",
    collaboration = "LIGO Scientific, Virgo",
    title = "{Tests of General Relativity with GW170817}",
    eprint = "1811.00364",
    archivePrefix = "arXiv",
    primaryClass = "gr-qc",
    reportNumber = "LIGO-P1800059",
    doi = "10.1103/PhysRevLett.123.011102",
    journal = "Phys. Rev. Lett.",
    volume = "123",
    number = "1",
    pages = "011102",
    year = "2019"
}

@article{LIGOScientific:2021sio,
    author = "Abbott, R. and others",
    collaboration = "LIGO Scientific, VIRGO, KAGRA",
    title = "{Tests of General Relativity with GWTC-3}",
    eprint = "2112.06861",
    archivePrefix = "arXiv",
    primaryClass = "gr-qc",
    reportNumber = "LIGO-P2100275",
    month = "12",
    year = "2021"
}

@article{Miller:2025eak,
    author = "Miller, Simona J. and Isi, Maximiliano and Chatziioannou, Katerina and Varma, Vijay and Hourihane, Sophie",
    title = "{Measuring spin precession from massive black hole binaries with gravitational waves: insights from time-domain signal morphology}",
    eprint = "2505.14573",
    archivePrefix = "arXiv",
    primaryClass = "gr-qc",
    reportNumber = "LIGO-P2500263",
    month = "5",
    year = "2025"
}

@article{Husa:2015iqa,
    author = {Husa, Sascha and Khan, Sebastian and Hannam, Mark and P{\"u}rrer, Michael and Ohme, Frank and Jim{\'e}nez Forteza, Xisco and Boh{\'e}, Alejandro},
    title = "{Frequency-domain gravitational waves from nonprecessing black-hole binaries. I. New numerical waveforms and anatomy of the signal}",
    eprint = "1508.07250",
    archivePrefix = "arXiv",
    primaryClass = "gr-qc",
    doi = "10.1103/PhysRevD.93.044006",
    journal = "Phys. Rev. D",
    volume = "93",
    number = "4",
    pages = "044006",
    year = "2016"
}

@article{Thomas:2025rje,
    author = "Thomas, Lucy M. and Chatziioannou, Katerina and Varma, Vijay and Field, Scott E.",
    title = "{Optimizing neural network surrogate models: Application to black hole merger remnants}",
    eprint = "2501.16462",
    archivePrefix = "arXiv",
    primaryClass = "gr-qc",
    reportNumber = "DCC:LIGO-P2400620",
    doi = "10.1103/PhysRevD.111.104029",
    journal = "Phys. Rev. D",
    volume = "111",
    number = "10",
    pages = "104029",
    year = "2025"
}

@article{Chattaraj:2022tay,
    author = "Chattaraj, Abhishek and RoyChowdhury, Tamal and Divyajyoti and Mishra, Chandra Kant and Gupta, Anshu",
    title = "{High accuracy post-Newtonian and numerical relativity comparisons involving higher modes for eccentric binary black holes and a dominant mode eccentric inspiral-merger-ringdown model}",
    eprint = "2204.02377",
    archivePrefix = "arXiv",
    primaryClass = "gr-qc",
    reportNumber = "LIGO-P2200106",
    doi = "10.1103/PhysRevD.106.124008",
    journal = "Phys. Rev. D",
    volume = "106",
    number = "12",
    pages = "124008",
    year = "2022"
}

@article{LIGOScientific:2014oec,
    author = "Aasi, J. and others",
    collaboration = "LIGO Scientific, VIRGO, NINJA-2",
    title = "{The NINJA-2 project: Detecting and characterizing gravitational waveforms modelled using numerical binary black hole simulations}",
    eprint = "1401.0939",
    archivePrefix = "arXiv",
    primaryClass = "gr-qc",
    reportNumber = "LIGO-P1300199",
    doi = "10.1088/0264-9381/31/11/115004",
    journal = "Class. Quant. Grav.",
    volume = "31",
    pages = "115004",
    year = "2014"
}

@article{Henry:2022dzx,
    author = "Henry, Quentin and Marsat, Sylvain and Khalil, Mohammed",
    title = "{Spin contributions to the gravitational-waveform modes for spin-aligned binaries at the 3.5PN order}",
    eprint = "2209.00374",
    archivePrefix = "arXiv",
    primaryClass = "gr-qc",
    doi = "10.1103/PhysRevD.106.124018",
    journal = "Phys. Rev. D",
    volume = "106",
    number = "12",
    pages = "124018",
    year = "2022"
}

@article{Kidder:2007rt,
    author = "Kidder, Lawrence E.",
    title = "{Using full information when computing modes of post-Newtonian waveforms from inspiralling compact binaries in circular orbit}",
    eprint = "0710.0614",
    archivePrefix = "arXiv",
    primaryClass = "gr-qc",
    doi = "10.1103/PhysRevD.77.044016",
    journal = "Phys. Rev. D",
    volume = "77",
    pages = "044016",
    year = "2008"
}

@article{Aylott:2009ya,
    author = "Aylott, Benjamin and others",
    title = "{Testing gravitational-wave searches with numerical relativity waveforms: Results from the first Numerical INJection Analysis (NINJA) project}",
    eprint = "0901.4399",
    archivePrefix = "arXiv",
    primaryClass = "gr-qc",
    doi = "10.1088/0264-9381/26/16/165008",
    journal = "Class. Quant. Grav.",
    volume = "26",
    pages = "165008",
    year = "2009"
}

@article{Lovelace:2016uwp,
    author = "Lovelace, Geoffrey and others",
    title = "{Modeling the source of GW150914 with targeted numerical-relativity simulations}",
    eprint = "1607.05377",
    archivePrefix = "arXiv",
    primaryClass = "gr-qc",
    doi = "10.1088/0264-9381/33/24/244002",
    journal = "Class. Quant. Grav.",
    volume = "33",
    number = "24",
    pages = "244002",
    year = "2016"
}

@article{Wang:2024iyj,
    author = "Wang, Zun and Zhao, Junjie and Cao, Zhoujian",
    title = "{Accuracy of numerical relativity waveforms with respect to space-based gravitational wave detectors}",
    eprint = "2401.15331",
    archivePrefix = "arXiv",
    primaryClass = "gr-qc",
    doi = "10.1088/1572-9494/ad1824",
    journal = "Commun. Theor. Phys.",
    volume = "76",
    number = "1",
    pages = "015403",
    year = "2024"
}

@article{Zlochower:2012fk,
    author = "Zlochower, Yosef and Ponce, Marcelo and Lousto, Carlos O.",
    title = "{Accuracy Issues for Numerical Waveforms}",
    eprint = "1208.5494",
    archivePrefix = "arXiv",
    primaryClass = "gr-qc",
    doi = "10.1103/PhysRevD.86.104056",
    journal = "Phys. Rev. D",
    volume = "86",
    pages = "104056",
    year = "2012"
}

@article{Okounkova:2020rqw,
    author = "Okounkova, Maria",
    title = "{Numerical relativity simulation of GW150914 in Einstein dilaton Gauss-Bonnet gravity}",
    eprint = "2001.03571",
    archivePrefix = "arXiv",
    primaryClass = "gr-qc",
    doi = "10.1103/PhysRevD.102.084046",
    journal = "Phys. Rev. D",
    volume = "102",
    number = "8",
    pages = "084046",
    year = "2020"
}

@article{Okounkova:2019zjf,
    author = "Okounkova, Maria and Stein, Leo C. and Moxon, Jordan and Scheel, Mark A. and Teukolsky, Saul A.",
    title = "{Numerical relativity simulation of GW150914 beyond general relativity}",
    eprint = "1911.02588",
    archivePrefix = "arXiv",
    primaryClass = "gr-qc",
    doi = "10.1103/PhysRevD.101.104016",
    journal = "Phys. Rev. D",
    volume = "101",
    number = "10",
    pages = "104016",
    year = "2020"
}

@article{Lara:2025kzj,
    author = "Lara, Guillermo and others",
    title = "{Signatures from metastable oppositely-charged black hole binaries in scalar Gauss-Bonnet gravity}",
    eprint = "2505.14785",
    archivePrefix = "arXiv",
    primaryClass = "gr-qc",
    month = "5",
    year = "2025"
}

@article{Gamba:2021ydi,
    author = "Gamba, Rossella and Ak{\c{c}}ay, Sarp and Bernuzzi, Sebastiano and Williams, Jake",
    title = "{Effective-one-body waveforms for precessing coalescing compact binaries with post-Newtonian twist}",
    eprint = "2111.03675",
    archivePrefix = "arXiv",
    primaryClass = "gr-qc",
    doi = "10.1103/PhysRevD.106.024020",
    journal = "Phys. Rev. D",
    volume = "106",
    number = "2",
    pages = "024020",
    year = "2022"
}

@article{Ramos-Buades:2023ehm,
    author = "Ramos-Buades, Antoni and Buonanno, Alessandra and Estell{\'e}s, H{\'e}ctor and Khalil, Mohammed and Mihaylov, Deyan P. and Ossokine, Serguei and Pompili, Lorenzo and Shiferaw, Mahlet",
    title = "{Next generation of accurate and efficient multipolar precessing-spin effective-one-body waveforms for binary black holes}",
    eprint = "2303.18046",
    archivePrefix = "arXiv",
    primaryClass = "gr-qc",
    doi = "10.1103/PhysRevD.108.124037",
    journal = "Phys. Rev. D",
    volume = "108",
    number = "12",
    pages = "124037",
    year = "2023"
}

@article{Cotesta:2018fcv,
    author = "Cotesta, Roberto and Buonanno, Alessandra and Boh{\'e}, Alejandro and Taracchini, Andrea and Hinder, Ian and Ossokine, Serguei",
    title = "{Enriching the Symphony of Gravitational Waves from Binary Black Holes by Tuning Higher Harmonics}",
    eprint = "1803.10701",
    archivePrefix = "arXiv",
    primaryClass = "gr-qc",
    doi = "10.1103/PhysRevD.98.084028",
    journal = "Phys. Rev. D",
    volume = "98",
    number = "8",
    pages = "084028",
    year = "2018"
}

@article{Hannam:2007ik,
    author = "Hannam, Mark and Husa, Sascha and Sperhake, Ulrich and Bruegmann, Bernd and Gonzalez, Jose A.",
    title = "{Where post-Newtonian and numerical-relativity waveforms meet}",
    eprint = "0706.1305",
    archivePrefix = "arXiv",
    primaryClass = "gr-qc",
    doi = "10.1103/PhysRevD.77.044020",
    journal = "Phys. Rev. D",
    volume = "77",
    pages = "044020",
    year = "2008"
}

@article{Okounkova:2019dfo,
    author = "Okounkova, Maria and Stein, Leo C. and Scheel, Mark A. and Teukolsky, Saul A.",
    title = "{Numerical binary black hole collisions in dynamical Chern-Simons gravity}",
    eprint = "1906.08789",
    archivePrefix = "arXiv",
    primaryClass = "gr-qc",
    doi = "10.1103/PhysRevD.100.104026",
    journal = "Phys. Rev. D",
    volume = "100",
    number = "10",
    pages = "104026",
    year = "2019"
}

@article{Gamboa:2024hli,
    author = "Gamboa, Aldo and others",
    title = "{Accurate waveforms for eccentric, aligned-spin binary black holes: The multipolar effective-one-body model seobnrv5ehm}",
    eprint = "2412.12823",
    archivePrefix = "arXiv",
    primaryClass = "gr-qc",
    doi = "10.1103/jxrc-z298",
    journal = "Phys. Rev. D",
    volume = "112",
    number = "4",
    pages = "044038",
    year = "2025"
}

@article{Paul:2024ujx,
    author = "Paul, Kaushik and Maurya, Akash and Henry, Quentin and Sharma, Kartikey and Satheesh, Pranav and Divyajyoti and Kumar, Prayush and Mishra, Chandra Kant",
    title = "{Eccentric, spinning, inspiral-merger-ringdown waveform model with higher modes for the detection and characterization of binary black holes}",
    eprint = "2409.13866",
    archivePrefix = "arXiv",
    primaryClass = "gr-qc",
    doi = "10.1103/PhysRevD.111.084074",
    journal = "Phys. Rev. D",
    volume = "111",
    number = "8",
    pages = "084074",
    year = "2025"
}

@article{Hinder:2017sxy,
    author = "Hinder, Ian and Kidder, Lawrence E. and Pfeiffer, Harald P.",
    title = "{Eccentric binary black hole inspiral-merger-ringdown gravitational waveform model from numerical relativity and post-Newtonian theory}",
    eprint = "1709.02007",
    archivePrefix = "arXiv",
    primaryClass = "gr-qc",
    doi = "10.1103/PhysRevD.98.044015",
    journal = "Phys. Rev. D",
    volume = "98",
    number = "4",
    pages = "044015",
    year = "2018"
}

@article{Huerta:2016rwp,
    author = "Huerta, E. A. and others",
    title = "{Complete waveform model for compact binaries on eccentric orbits}",
    eprint = "1609.05933",
    archivePrefix = "arXiv",
    primaryClass = "gr-qc",
    doi = "10.1103/PhysRevD.95.024038",
    journal = "Phys. Rev. D",
    volume = "95",
    number = "2",
    pages = "024038",
    year = "2017"
}

@article{Ramos-Buades:2019uvh,
    author = "Ramos-Buades, Antoni and Husa, Sascha and Pratten, Geraint and Estell{\'e}s, H{\'e}ctor and Garc{\'\i}a-Quir{\'o}s, Cecilio and Mateu-Lucena, Maite and Colleoni, Marta and Jaume, Rafel",
    title = "{First survey of spinning eccentric black hole mergers: Numerical relativity simulations, hybrid waveforms, and parameter estimation}",
    eprint = "1909.11011",
    archivePrefix = "arXiv",
    primaryClass = "gr-qc",
    doi = "10.1103/PhysRevD.101.083015",
    journal = "Phys. Rev. D",
    volume = "101",
    number = "8",
    pages = "083015",
    year = "2020"
}

@article{LIGOScientific:2025epi,
    author = "Abac, A. G. and others",
    collaboration = "LIGO Scientific, KAGRA, Virgo",
    title = "{GW250114: Testing Hawking{\textquoteright}s Area Law and the Kerr Nature of Black Holes}",
    eprint = "2509.08054",
    archivePrefix = "arXiv",
    primaryClass = "gr-qc",
    reportNumber = "LIGO-P2500421",
    doi = "10.1103/kw5g-d732",
    journal = "Phys. Rev. Lett.",
    volume = "135",
    number = "11",
    pages = "111403",
    year = "2025"
}

@article{Mould:2021xst,
    author = "Mould, Matthew and Gerosa, Davide",
    title = "{Gravitational-wave population inference at past time infinity}",
    eprint = "2110.05507",
    archivePrefix = "arXiv",
    primaryClass = "astro-ph.HE",
    doi = "10.1103/PhysRevD.105.024076",
    journal = "Phys. Rev. D",
    volume = "105",
    number = "2",
    pages = "024076",
    year = "2022"
}

@ARTICLE{1980ApJ...239..292D,
       author = {{Detweiler}, S.},
        title = "{Black holes and gravitational waves. III - The resonant frequencies of rotating holes}",
      journal = {\apj},
         year = 1980,
        month = jul,
       volume = {239},
        pages = {292-295},
          doi = {10.1086/158109},
       adsurl = {https://ui.adsabs.harvard.edu/abs/1980ApJ...239..292D}
}

@ARTICLE{1973ApJ...185..635T,
       author = {{Teukolsky}, Saul A.},
        title = "{Perturbations of a Rotating Black Hole. I. Fundamental Equations for Gravitational, Electromagnetic, and Neutrino-Field Perturbations}",
      journal = {\apj},
         year = 1973,
        month = oct,
       volume = {185},
        pages = {635-648},
          doi = {10.1086/152444},
       adsurl = {https://ui.adsabs.harvard.edu/abs/1973ApJ...185..635T}
}

@article{Schmidt_2015,
   title={Towards models of gravitational waveforms from generic binaries: II. Modelling precession effects with a single effective precession parameter},
   volume={91},
   ISSN={1550-2368},
   url={http://dx.doi.org/10.1103/PhysRevD.91.024043},
   DOI={10.1103/physrevd.91.024043},
   number={2},
   journal={Physical Review D},
   publisher={American Physical Society (APS)},
   author={Schmidt, Patricia and Ohme, Frank and Hannam, Mark},
   year={2015},
   month=jan }

@article{Isi:2019aib,
    author = "Isi, Maximiliano and Giesler, Matthew and Farr, Will M. and Scheel, Mark A. and Teukolsky, Saul A.",
    title = "{Testing the no-hair theorem with GW150914}",
    eprint = "1905.00869",
    archivePrefix = "arXiv",
    primaryClass = "gr-qc",
    reportNumber = "LIGO-P1900135",
    doi = "10.1103/PhysRevLett.123.111102",
    journal = "Phys. Rev. Lett.",
    volume = "123",
    number = "11",
    pages = "111102",
    year = "2019"
}

@article{Colleoni:2024knd,
    author = "Colleoni, Marta and Vidal, Felip A. Ramis and Garc{\'\i}a-Quir{\'o}s, Cecilio and Ak{\c{c}}ay, Sarp and Bera, Sayantani",
    title = "{Fast frequency-domain gravitational waveforms for precessing binaries with a new twist}",
    eprint = "2412.16721",
    archivePrefix = "arXiv",
    primaryClass = "gr-qc",
    doi = "10.1103/PhysRevD.111.104019",
    journal = "Phys. Rev. D",
    volume = "111",
    number = "10",
    pages = "104019",
    year = "2025"
}

@article{Cotesta:2022pci,
    author = "Cotesta, Roberto and Carullo, Gregorio and Berti, Emanuele and Cardoso, Vitor",
    title = "{Analysis of Ringdown Overtones in GW150914}",
    eprint = "2201.00822",
    archivePrefix = "arXiv",
    primaryClass = "gr-qc",
    doi = "10.1103/PhysRevLett.129.111102",
    journal = "Phys. Rev. Lett.",
    volume = "129",
    number = "11",
    pages = "111102",
    year = "2022"
}

@article{Pratten:2020ceb,
    author = "Pratten, Geraint and others",
    title = "{Computationally efficient models for the dominant and subdominant harmonic modes of precessing binary black holes}",
    eprint = "2004.06503",
    archivePrefix = "arXiv",
    primaryClass = "gr-qc",
    doi = "10.1103/PhysRevD.103.104056",
    journal = "Phys. Rev. D",
    volume = "103",
    number = "10",
    pages = "104056",
    year = "2021"
}

@article{Branchesi:2023mws,
    author = "Branchesi, Marica and others",
    title = "{Science with the Einstein Telescope: a comparison of different designs}",
    eprint = "2303.15923",
    archivePrefix = "arXiv",
    primaryClass = "gr-qc",
    reportNumber = "ET-0084A-23",
    doi = "10.1088/1475-7516/2023/07/068",
    journal = "JCAP",
    volume = "07",
    pages = "068",
    year = "2023"
}

@article{Baird:2012cu,
    author = "Baird, Emily and Fairhurst, Stephen and Hannam, Mark and Murphy, Patricia",
    title = "{Degeneracy between mass and spin in black-hole-binary waveforms}",
    eprint = "1211.0546",
    archivePrefix = "arXiv",
    primaryClass = "gr-qc",
    doi = "10.1103/PhysRevD.87.024035",
    journal = "Phys. Rev. D",
    volume = "87",
    number = "2",
    pages = "024035",
    year = "2013"
}

@article{Mitman:2021xkq,
    author = "Mitman, Keefe and others",
    title = "{Fixing the BMS frame of numerical relativity waveforms}",
    eprint = "2105.02300",
    archivePrefix = "arXiv",
    primaryClass = "gr-qc",
    doi = "10.1103/PhysRevD.104.024051",
    journal = "Phys. Rev. D",
    volume = "104",
    number = "2",
    pages = "024051",
    year = "2021"
}

@article{Ajith:2009bn,
    author = "Ajith, P. and others",
    title = "{Inspiral-merger-ringdown waveforms for black-hole binaries with non-precessing spins}",
    eprint = "0909.2867",
    archivePrefix = "arXiv",
    primaryClass = "gr-qc",
    doi = "10.1103/PhysRevLett.106.241101",
    journal = "Phys. Rev. Lett.",
    volume = "106",
    pages = "241101",
    year = "2011"
}

@article{Payne:2022spz,
    author = "Payne, Ethan and Hourihane, Sophie and Golomb, Jacob and Udall, Rhiannon and Udall, Richard and Davis, Derek and Chatziioannou, Katerina",
    title = "{Curious case of GW200129: Interplay between spin-precession inference and data-quality issues}",
    eprint = "2206.11932",
    archivePrefix = "arXiv",
    primaryClass = "gr-qc",
    reportNumber = "LIGO DCC: P2200185",
    doi = "10.1103/PhysRevD.106.104017",
    journal = "Phys. Rev. D",
    volume = "106",
    number = "10",
    pages = "104017",
    year = "2022"
}

@article{Buscicchio:2024asl,
    author = "Buscicchio, R. and Torrado, J. and Caprini, C. and Nardini, G. and Karnesis, N. and Pieroni, M. and Sesana, A.",
    title = "{Stellar-mass black-hole binaries in LISA: characteristics and complementarity with current-generation interferometers}",
    eprint = "2410.18171",
    archivePrefix = "arXiv",
    primaryClass = "astro-ph.HE",
    reportNumber = "CERN-TH-2024-173",
    doi = "10.1088/1475-7516/2025/01/084",
    journal = "JCAP",
    volume = "01",
    pages = "084",
    year = "2025"
}

@article{Toubiana:2022vpp,
    author = "Toubiana, Alexandre and Babak, Stanislav and Marsat, Sylvain and Ossokine, Sergei",
    title = "{Detectability and parameter estimation of GWTC-3 events with LISA}",
    eprint = "2206.12439",
    archivePrefix = "arXiv",
    primaryClass = "gr-qc",
    doi = "10.1103/PhysRevD.106.104034",
    journal = "Phys. Rev. D",
    volume = "106",
    number = "10",
    pages = "104034",
    year = "2022"
}

@article{Boyle:2015nqa,
    author = "Boyle, Michael",
    title = "{Transformations of asymptotic gravitational-wave data}",
    eprint = "1509.00862",
    archivePrefix = "arXiv",
    primaryClass = "gr-qc",
    doi = "10.1103/PhysRevD.93.084031",
    journal = "Phys. Rev. D",
    volume = "93",
    number = "8",
    pages = "084031",
    year = "2016"
}

@article{DaRe:2025glj,
    author = "Da Re, Guido and others",
    title = "{Modeling the BMS transformation induced by a binary black hole merger}",
    eprint = "2503.09569",
    archivePrefix = "arXiv",
    primaryClass = "gr-qc",
    doi = "10.1103/PhysRevD.111.124019",
    journal = "Phys. Rev. D",
    volume = "111",
    number = "12",
    pages = "124019",
    year = "2025"
}

@article{Campanelli:2006uy,
    author = "Campanelli, Manuela and Lousto, C. O. and Zlochower, Y.",
    title = "{Spinning-black-hole binaries: The orbital hang up}",
    eprint = "gr-qc/0604012",
    archivePrefix = "arXiv",
    doi = "10.1103/PhysRevD.74.041501",
    journal = "Phys. Rev. D",
    volume = "74",
    pages = "041501",
    year = "2006"
}

@article{Miller:2023ncs,
    author = "Miller, Simona J. and Isi, Maximiliano and Chatziioannou, Katerina and Varma, Vijay and Mandel, Ilya",
    title = "{GW190521: Tracing imprints of spin-precession on the most massive black hole binary}",
    eprint = "2310.01544",
    archivePrefix = "arXiv",
    primaryClass = "astro-ph.HE",
    reportNumber = "LIGO-P2300329",
    doi = "10.1103/PhysRevD.109.024024",
    journal = "Phys. Rev. D",
    volume = "109",
    number = "2",
    pages = "024024",
    year = "2024"
}

@misc{reitze2019cosmicexploreruscontribution,
      title={Cosmic Explorer: The U.S. Contribution to Gravitational-Wave Astronomy beyond LIGO}, 
      author={David Reitze and Rana X Adhikari and Stefan Ballmer and Barry Barish and Lisa Barsotti and GariLynn Billingsley and Duncan A. Brown and Yanbei Chen and Dennis Coyne and Robert Eisenstein and Matthew Evans and Peter Fritschel and Evan D. Hall and Albert Lazzarini and Geoffrey Lovelace and Jocelyn Read and B. S. Sathyaprakash and David Shoemaker and Joshua Smith and Calum Torrie and Salvatore Vitale and Rainer Weiss and Christopher Wipf and Michael Zucker},
      year={2019},
      eprint={1907.04833},
      archivePrefix={arXiv},
      primaryClass={astro-ph.IM},
      url={https://arxiv.org/abs/1907.04833}, 
}

@misc{amaroseoane2017laserinterferometerspaceantenna,
      title={Laser Interferometer Space Antenna}, 
      author={Pau Amaro-Seoane and Heather Audley and Stanislav Babak and John Baker and Enrico Barausse and Peter Bender and Emanuele Berti and Pierre Binetruy and Michael Born and Daniele Bortoluzzi and Jordan Camp and Chiara Caprini and Vitor Cardoso and Monica Colpi and John Conklin and Neil Cornish and Curt Cutler and Karsten Danzmann and Rita Dolesi and Luigi Ferraioli and Valerio Ferroni and Ewan Fitzsimons and Jonathan Gair and Lluis Gesa Bote and Domenico Giardini and Ferran Gibert and Catia Grimani and Hubert Halloin and Gerhard Heinzel and Thomas Hertog and Martin Hewitson and Kelly Holley-Bockelmann and Daniel Hollington and Mauro Hueller and Henri Inchauspe and Philippe Jetzer and Nikos Karnesis and Christian Killow and Antoine Klein and Bill Klipstein and Natalia Korsakova and Shane L Larson and Jeffrey Livas and Ivan Lloro and Nary Man and Davor Mance and Joseph Martino and Ignacio Mateos and Kirk McKenzie and Sean T McWilliams and Cole Miller and Guido Mueller and Germano Nardini and Gijs Nelemans and Miquel Nofrarias and Antoine Petiteau and Paolo Pivato and Eric Plagnol and Ed Porter and Jens Reiche and David Robertson and Norna Robertson and Elena Rossi and Giuliana Russano and Bernard Schutz and Alberto Sesana and David Shoemaker and Jacob Slutsky and Carlos F. Sopuerta and Tim Sumner and Nicola Tamanini and Ira Thorpe and Michael Troebs and Michele Vallisneri and Alberto Vecchio and Daniele Vetrugno and Stefano Vitale and Marta Volonteri and Gudrun Wanner and Harry Ward and Peter Wass and William Weber and John Ziemer and Peter Zweifel},
      year={2017},
      eprint={1702.00786},
      archivePrefix={arXiv},
      primaryClass={astro-ph.IM},
      url={https://arxiv.org/abs/1702.00786}, 
}

@article{Ossokine_2015,
   title={Improvements to the construction of binary black hole initial data},
   volume={32},
   ISSN={1361-6382},
   url={http://dx.doi.org/10.1088/0264-9381/32/24/245010},
   DOI={10.1088/0264-9381/32/24/245010},
   number={24},
   journal={Classical and Quantum Gravity},
   publisher={IOP Publishing},
   author={Ossokine, Serguei and Foucart, Francois and Pfeiffer, Harald P and Boyle, Michael and Szilágyi, Béla},
   year={2015},
   month=dec, pages={245010} }

@article{Boyle:2009,
  title = {Extrapolating gravitational-wave data from numerical simulations},
  author = {Boyle, Michael and Mrou\'e, Abdul H.},
  journal = {Phys. Rev. D},
  volume = {80},
  issue = {12},
  pages = {124045},
  numpages = {14},
  year = {2009},
  month = {Dec},
  publisher = {American Physical Society},
  doi = {10.1103/PhysRevD.80.124045},
  url = {https://link.aps.org/doi/10.1103/PhysRevD.80.124045}
}

@article{Varma_2019,
   title={Surrogate models for precessing binary black hole simulations with unequal masses},
   volume={1},
   ISSN={2643-1564},
   url={http://dx.doi.org/10.1103/PhysRevResearch.1.033015},
   DOI={10.1103/physrevresearch.1.033015},
   number={3},
   journal={Physical Review Research},
   publisher={American Physical Society (APS)},
   author={Varma, Vijay and Field, Scott E. and Scheel, Mark A. and Blackman, Jonathan and Gerosa, Davide and Stein, Leo C. and Kidder, Lawrence E. and Pfeiffer, Harald P.},
   year={2019},
   month=oct }

@article{Varma:2021csh,
    author = "Varma, Vijay and Isi, Maximiliano and Biscoveanu, Sylvia and Farr, Will M. and Vitale, Salvatore",
    title = "{Measuring binary black hole orbital-plane spin orientations}",
    eprint = "2107.09692",
    archivePrefix = "arXiv",
    primaryClass = "astro-ph.HE",
    doi = "10.1103/PhysRevD.105.024045",
    journal = "Phys. Rev. D",
    volume = "105",
    number = "2",
    pages = "024045",
    year = "2022"
}

@article{Varma:2019,
   title={Surrogate model of hybridized numerical relativity binary black hole waveforms},
   volume={99},
   ISSN={2470-0029},
   url={http://dx.doi.org/10.1103/PhysRevD.99.064045},
   DOI={10.1103/physrevd.99.064045},
   number={6},
   journal={Physical Review D},
   publisher={American Physical Society (APS)},
   author={Varma, Vijay and Field, Scott E. and Scheel, Mark A. and Blackman, Jonathan and Kidder, Lawrence E. and Pfeiffer, Harald P.},
   year={2019},
   month=mar }

@article{Lovelace:2011,
  title = {Simulating merging binary black holes with nearly extremal spins},
  author = {Lovelace, Geoffrey and Scheel, Mark A. and Szil\'agyi, B\'ela},
  journal = {Phys. Rev. D},
  volume = {83},
  issue = {2},
  pages = {024010},
  numpages = {5},
  year = {2011},
  month = {Jan},
  publisher = {American Physical Society},
  doi = {10.1103/PhysRevD.83.024010},
  url = {https://link.aps.org/doi/10.1103/PhysRevD.83.024010}
}

@article{Szilagyi:2014,
   title={Key elements of robustness in binary black hole evolutions using spectral methods},
   volume={23},
   ISSN={1793-6594},
   url={http://dx.doi.org/10.1142/S0218271814300146},
   DOI={10.1142/s0218271814300146},
   number={07},
   journal={International Journal of Modern Physics D},
   publisher={World Scientific Pub Co Pte Lt},
   author={Szilágyi, Béla},
   year={2014},
   month=jun, pages={1430014} }

@article{Pretto:2024dvx,
    author = "Pretto, Isabella G. and Scheel, Mark A. and Teukolsky, Saul A.",
    title = "{Automated determination of the end time of junk radiation in binary black hole simulations}",
    eprint = "2407.20470",
    archivePrefix = "arXiv",
    primaryClass = "gr-qc",
    month = "7",
    year = "2024"
}

@article{Lovelace_2011,
   title={Simulating merging binary black holes with nearly extremal spins},
   volume={83},
   ISSN={1550-2368},
   url={http://dx.doi.org/10.1103/PhysRevD.83.024010},
   DOI={10.1103/physrevd.83.024010},
   number={2},
   journal={Physical Review D},
   publisher={American Physical Society (APS)},
   author={Lovelace, Geoffrey and Scheel, Mark A. and Szilágyi, Béla},
   year={2011},
   month=jan }

@misc{purrer2019readyliesahead,
      title={Ready for what lies ahead? -- Gravitational waveform accuracy requirements for future ground based detectors}, 
      author={Michael Pürrer and Carl-Johan Haster},
      year={2019},
      eprint={1912.10055},
      archivePrefix={arXiv},
      primaryClass={gr-qc},
      url={https://arxiv.org/abs/1912.10055}, 
}

@article{Knapp:2024yww,
    author = "Knapp, Taylor and Chatziioannou, Katerina and Pfeiffer, Harald and Scheel, Mark A. and Kidder, Lawrence E.",
    title = "{Parameter control for eccentric, precessing binary black hole simulations with SpEC}",
    eprint = "2410.02997",
    archivePrefix = "arXiv",
    primaryClass = "gr-qc",
    doi = "10.1103/PhysRevD.111.024003",
    journal = "Phys. Rev. D",
    volume = "111",
    number = "2",
    pages = "024003",
    year = "2025"
}

@article{Habib:2024soh,
    author = "Habib, Sarah and Scheel, Mark and Teukolsky, Saul",
    title = "{Eccentricity Reduction for Quasicircular Binary Evolutions}",
    eprint = "2410.05531",
    archivePrefix = "arXiv",
    primaryClass = "gr-qc",
    month = "10",
    year = "2024"
}

@article{VIRGO:2014yos,
    author = "Acernese, F. and others",
    collaboration = "VIRGO",
    title = "{Advanced Virgo: a second-generation interferometric gravitational wave detector}",
    eprint = "1408.3978",
    archivePrefix = "arXiv",
    primaryClass = "gr-qc",
    doi = "10.1088/0264-9381/32/2/024001",
    journal = "Class. Quant. Grav.",
    volume = "32",
    number = "2",
    pages = "024001",
    year = "2015"
}

@article{Szil_gyi_2014,
   title={Key elements of robustness in binary black hole evolutions using spectral methods},
   volume={23},
   ISSN={1793-6594},
   url={http://dx.doi.org/10.1142/S0218271814300146},
   DOI={10.1142/s0218271814300146},
   number={07},
   journal={International Journal of Modern Physics D},
   publisher={World Scientific Pub Co Pte Lt},
   author={Szilágyi, Béla},
   year={2014},
   month=jun, pages={1430014} }

@article{Chu_2016,
   title={On the accuracy and precision of numerical waveforms: effect of waveform extraction methodology},
   volume={33},
   ISSN={1361-6382},
   url={http://dx.doi.org/10.1088/0264-9381/33/16/165001},
   DOI={10.1088/0264-9381/33/16/165001},
   number={16},
   journal={Classical and Quantum Gravity},
   publisher={IOP Publishing},
   author={Chu, Tony and Fong, Heather and Kumar, Prayush and Pfeiffer, Harald P and Boyle, Michael and Hemberger, Daniel A and Kidder, Lawrence E and Scheel, Mark A and Szilagyi, Bela},
   year={2016},
   month=jul, pages={165001} }

@misc{ferguson2023secondmayacatalogbinary,
      title={Second MAYA Catalog of Binary Black Hole Numerical Relativity Waveforms}, 
      author={Deborah Ferguson and Evelyn Allsup and Surendra Anne and Galina Bouyer and Miguel Gracia-Linares and Hector Iglesias and Aasim Jan and Pablo Laguna and Jacob Lange and Erick Martinez and Filippo Meoni and Ryan Nowicki and Deirdre Shoemaker and Blake Steadham and Max L. Trostel and Bing-Jyun Tsao and Finny Valorz},
      year={2023},
      eprint={2309.00262},
      archivePrefix={arXiv},
      primaryClass={gr-qc},
      url={https://arxiv.org/abs/2309.00262}, 
}

@book{Kopriva_2009,
  title     = "Implementing Spectral Methods for Partial Differential Equations",
  author    = "Kopriva, David A.",
  year      = 2009,
  publisher = "Springer Dordrecht"
}

@article{Hinder:2013oqa,
    author = "Hinder, Ian and others",
    title = "{Error-analysis and comparison to analytical models of numerical waveforms produced by the NRAR Collaboration}",
    eprint = "1307.5307",
    archivePrefix = "arXiv",
    primaryClass = "gr-qc",
    reportNumber = "AEI-2013-223",
    doi = "10.1088/0264-9381/31/2/025012",
    journal = "Class. Quant. Grav.",
    volume = "31",
    pages = "025012",
    year = "2014"
}

@article{Aylott:2009tn,
    author = "Aylott, Benjamin and others",
    editor = "Sutton, Patrick and Shoemaker, Deirdre",
    title = "{Status of NINJA: The Numerical INJection Analysis project}",
    eprint = "0905.4227",
    archivePrefix = "arXiv",
    primaryClass = "gr-qc",
    doi = "10.1088/0264-9381/26/11/114008",
    journal = "Class. Quant. Grav.",
    volume = "26",
    pages = "114008",
    year = "2009"
}

@article{Huerta_2019,
   title={Physics of eccentric binary black hole mergers: A numerical relativity perspective},
   volume={100},
   ISSN={2470-0029},
   url={http://dx.doi.org/10.1103/PhysRevD.100.064003},
   DOI={10.1103/physrevd.100.064003},
   number={6},
   journal={Physical Review D},
   publisher={American Physical Society (APS)},
   author={Huerta, E. A. and Haas, Roland and Habib, Sarah and Gupta, Anushri and Rebei, Adam and Chavva, Vishnu and Johnson, Daniel and Rosofsky, Shawn and Wessel, Erik and Agarwal, Bhanu and Luo, Diyu and Ren, Wei},
   year={2019},
   month=sep }

@misc{hamilton2023catalogueprecessingblackholebinarynumericalrelativity,
      title={A catalogue of precessing black-hole-binary numerical-relativity simulations}, 
      author={Eleanor Hamilton and Edward Fauchon-Jones and Mark Hannam and Charlie Hoy and Chinmay Kalaghatgi and Lionel London and Jonathan E. Thompson and Dave Yeeles and Shrobana Ghosh and Sebastian Khan and Panagiota Kolitsidou and Alex Vano-Vinuales},
      year={2023},
      eprint={2303.05419},
      archivePrefix={arXiv},
      primaryClass={gr-qc},
      url={https://arxiv.org/abs/2303.05419}, 
}

@article{Rashti_2025,
   title={Binary black hole waveforms from high-resolution gr-athena++ simulations},
   volume={111},
   ISSN={2470-0029},
   url={http://dx.doi.org/10.1103/n5pz-qv3x},
   DOI={10.1103/n5pz-qv3x},
   number={10},
   journal={Physical Review D},
   publisher={American Physixcal Society (APS)},
   author={Rashti, Alireza and Gamba, Rossella and Chandra, Koustav and Radice, David and Daszuta, Boris and Cook, William and Bernuzzi, Sebastiano},
   year={2025},
   month=may }

@article{Healy_2020,
   title={Third RIT binary black hole simulations catalog},
   volume={102},
   ISSN={2470-0029},
   url={http://dx.doi.org/10.1103/PhysRevD.102.104018},
   DOI={10.1103/physrevd.102.104018},
   number={10},
   journal={Physical Review D},
   publisher={American Physical Society (APS)},
   author={Healy, James and Lousto, Carlos O.},
   year={2020},
   month=nov }

@article{LIGOScientific:2014pky,
    author = "Aasi, J. and others",
    collaboration = "LIGO Scientific",
    title = "{Advanced LIGO}",
    eprint = "1411.4547",
    archivePrefix = "arXiv",
    primaryClass = "gr-qc",
    doi = "10.1088/0264-9381/32/7/074001",
    journal = "Class. Quant. Grav.",
    volume = "32",
    pages = "074001",
    year = "2015"
}

@misc{thompson2025,
      title={On the use and interpretation of signal-model indistinguishability measures for gravitational-wave astronomy}, 
      author={Jonathan E. Thompson and Charlie Hoy and Edward Fauchon-Jones and Mark Hannam},
      year={2025},
      eprint={2506.10530},
      archivePrefix={arXiv},
      primaryClass={gr-qc},
      url={https://arxiv.org/abs/2506.10530}, 
}

@misc{ajith2011dataformatsnumericalrelativity,
      title={Data formats for numerical relativity waves}, 
      author={P. Ajith and M. Boyle and D. A. Brown and S. Fairhurst and M. Hannam and I. Hinder and S. Husa and B. Krishnan and R. A. Mercer and F. Ohme and C. D. Ott and J. S. Read and L. Santamaria and J. T. Whelan},
      year={2011},
      eprint={0709.0093},
      archivePrefix={arXiv},
      primaryClass={gr-qc},
      url={https://arxiv.org/abs/0709.0093}, 
}

@article{Mitman:2025tmj,
    author = "Mitman, Keefe and Stein, Leo C. and Boyle, Michael and Deppe, Nils and Kidder, Lawrence E. and Pfeiffer, Harald P. and Scheel, Mark A.",
    title = "{Length dependence of waveform mismatch: a caveat on waveform accuracy}",
    eprint = "2502.14025",
    archivePrefix = "arXiv",
    primaryClass = "gr-qc",
    month = "2",
    year = "2025"
}

@article{Zlochower_2012,
   title={Accuracy issues for numerical waveforms},
   volume={86},
   ISSN={1550-2368},
   url={http://dx.doi.org/10.1103/PhysRevD.86.104056},
   DOI={10.1103/physrevd.86.104056},
   number={10},
   journal={Physical Review D},
   publisher={American Physical Society (APS)},
   author={Zlochower, Yosef and Ponce, Marcelo and Lousto, Carlos O.},
   year={2012},
   month=nov }

@article{Williamson_2017,
   title={Systematic challenges for future gravitational wave measurements of precessing binary black holes},
   volume={96},
   ISSN={2470-0029},
   url={http://dx.doi.org/10.1103/PhysRevD.96.124041},
   DOI={10.1103/physrevd.96.124041},
   number={12},
   journal={Physical Review D},
   publisher={American Physical Society (APS)},
   author={Williamson, A. R. and Lange, J. and O’Shaughnessy, R. and Clark, J. A. and Kumar, Prayush and Calderón Bustillo, J. and Veitch, J.},
   year={2017},
   month=dec }

@article{Jan_2020,
   title={Assessing and marginalizing over compact binary coalescence waveform systematics with RIFT},
   volume={102},
   ISSN={2470-0029},
   url={http://dx.doi.org/10.1103/PhysRevD.102.124069},
   DOI={10.1103/physrevd.102.124069},
   number={12},
   journal={Physical Review D},
   publisher={American Physical Society (APS)},
   author={Jan, A. Z. and Yelikar, A. B. and Lange, J. and O’Shaughnessy, R.},
   year={2020},
   month=dec }

@misc{jan2024accuracylimitationsexistingnumerical,
      title={Accuracy limitations of existing numerical relativity waveforms on the data analysis of current and future ground-based detectors}, 
      author={Aasim Jan and Deborah Ferguson and Jacob Lange and Deirdre Shoemaker and Aaron Zimmerman},
      year={2024},
      eprint={2312.10241},
      archivePrefix={arXiv},
      primaryClass={gr-qc},
      url={https://arxiv.org/abs/2312.10241}, 
}

@article{Boh__2017,
   title={Improved effective-one-body model of spinning, nonprecessing binary black holes for the era of gravitational-wave astrophysics with advanced detectors},
   volume={95},
   ISSN={2470-0029},
   url={http://dx.doi.org/10.1103/PhysRevD.95.044028},
   DOI={10.1103/physrevd.95.044028},
   number={4},
   journal={Physical Review D},
   publisher={American Physical Society (APS)},
   author={Bohé, Alejandro and Shao, Lijing and Taracchini, Andrea and Buonanno, Alessandra and Babak, Stanislav and Harry, Ian W. and Hinder, Ian and Ossokine, Serguei and Pürrer, Michael and Raymond, Vivien and Chu, Tony and Fong, Heather and Kumar, Prayush and Pfeiffer, Harald P. and Boyle, Michael and Hemberger, Daniel A. and Kidder, Lawrence E. and Lovelace, Geoffrey and Scheel, Mark A. and Szilágyi, Béla},
   year={2017},
   month=feb }

@article{Buonanno_2007,
   title={Approaching faithful templates for nonspinning binary black holes using the effective-one-body approach},
   volume={76},
   ISSN={1550-2368},
   url={http://dx.doi.org/10.1103/PhysRevD.76.104049},
   DOI={10.1103/physrevd.76.104049},
   number={10},
   journal={Physical Review D},
   publisher={American Physical Society (APS)},
   author={Buonanno, Alessandra and Pan, Yi and Baker, John G. and Centrella, Joan and Kelly, Bernard J. and McWilliams, Sean T. and van Meter, James R.},
   year={2007},
   month=nov }

@article{Lousto_2023,
   title={Study of the intermediate mass ratio black hole binary merger up to 1000:1 with numerical relativity},
   volume={40},
   ISSN={1361-6382},
   url={http://dx.doi.org/10.1088/1361-6382/acc7ef},
   DOI={10.1088/1361-6382/acc7ef},
   number={9},
   journal={Classical and Quantum Gravity},
   publisher={IOP Publishing},
   author={Lousto, Carlos O and Healy, James},
   year={2023},
   month=apr, pages={09LT01} }

@article{Ferguson_2021,
   title={Assessing the readiness of numerical relativity for LISA and 3G detectors},
   volume={104},
   ISSN={2470-0029},
   url={http://dx.doi.org/10.1103/PhysRevD.104.044037},
   DOI={10.1103/physrevd.104.044037},
   number={4},
   journal={Physical Review D},
   publisher={American Physical Society (APS)},
   author={Ferguson, Deborah and Jani, Karan and Laguna, Pablo and Shoemaker, Deirdre},
   year={2021},
   month=aug }

@article{Mitman:2025hgy,
    author = "Mitman, Keefe and others",
    title = "{Probing the ringdown perturbation in binary black hole coalescences with an improved quasi-normal mode extraction algorithm}",
    eprint = "2503.09678",
    archivePrefix = "arXiv",
    primaryClass = "gr-qc",
    month = "3",
    year = "2025"
}

@article{Peters:1963ux,
    author = "Peters, P. C. and Mathews, J.",
    title = "{Gravitational radiation from point masses in a Keplerian orbit}",
    doi = "10.1103/PhysRev.131.435",
    journal = "Phys. Rev.",
    volume = "131",
    pages = "435--439",
    year = "1963"
}

@software{boyle2025scri,
  author       = {Boyle, Michael and Iozzo, Daniel and Stein, Leo and Khairnar, Anuj and Rüter, Hannes and Scheel, Mark and Varma, Vijay and Mitman, Keefe},
  title        = {scri (v2024.0.1)},
  year         = {2025},
  publisher    = {Zenodo},
  doi          = {10.5281/zenodo.15237110},
  url          = {https://doi.org/10.5281/zenodo.15237110},
}

@software{boyle2025sxs,
  doi = {10.5281/ZENODO.18765418},
  url = {https://zenodo.org/doi/10.5281/zenodo.18765418},
  author = {Boyle, Michael and Mitman, Keefe and Scheel, Mark and Stein, Leo},
  title = {The sxs package},
  publisher = {Zenodo},
  year = {2026},
  copyright = {MIT License}
}

@article{Scheel:2025jct,
    author = "Scheel, Mark A. and others",
    title = "{The SXS Collaboration's third catalog of binary black hole simulations}",
    eprint = "2505.13378",
    archivePrefix = "arXiv",
    primaryClass = "gr-qc",
    month = "5",
    year = "2025"
}

@misc{nagar_2023,
      title={Analytic systematics in next-generation of effective-one-body gravitational waveform models for future observations}, 
      author={Alessandro Nagar and Piero Rettegno and Rossella Gamba and Simone Albanesi and Angelica Albertini and Sebastiano Bernuzzi},
      year={2023},
      eprint={2304.09662},
      archivePrefix={arXiv},
      primaryClass={gr-qc},
      url={https://arxiv.org/abs/2304.09662}, 
}

@article{Biscoveanu:2021nvg,
    author = "Biscoveanu, Sylvia and Isi, Maximiliano and Varma, Vijay and Vitale, Salvatore",
    title = "{Measuring the spins of heavy binary black holes}",
    eprint = "2106.06492",
    archivePrefix = "arXiv",
    primaryClass = "gr-qc",
    reportNumber = "LIGO document number P2100204",
    doi = "10.1103/PhysRevD.104.103018",
    journal = "Phys. Rev. D",
    volume = "104",
    number = "10",
    pages = "103018",
    year = "2021"
}

@article{Nagar_2018,
   title={Time-domain effective-one-body gravitational waveforms for coalescing compact binaries with nonprecessing spins, tides, and self-spin effects},
   volume={98},
   ISSN={2470-0029},
   url={http://dx.doi.org/10.1103/PhysRevD.98.104052},
   DOI={10.1103/physrevd.98.104052},
   number={10},
   journal={Physical Review D},
   publisher={American Physical Society (APS)},
   author={Nagar, Alessandro and Bernuzzi, Sebastiano and Del Pozzo, Walter and Riemenschneider, Gunnar and Akcay, Sarp and Carullo, Gregorio and Fleig, Philipp and Babak, Stanislav and Tsang, Ka Wa and Colleoni, Marta and Messina, Francesco and Pratten, Geraint and Radice, David and Rettegno, Piero and Agathos, Michalis and Fauchon-Jones, Edward and Hannam, Mark and Husa, Sascha and Dietrich, Tim and Cerdá-Duran, Pablo and Font, José A. and Pannarale, Francesco and Schmidt, Patricia and Damour, Thibault},
   year={2018},
   month=nov }

@misc{pompili_2023,
      title={Laying the foundation of the effective-one-body waveform models SEOBNRv5: improved accuracy and efficiency for spinning non-precessing binary black holes}, 
      author={Lorenzo Pompili and Alessandra Buonanno and Héctor Estellés and Mohammed Khalil and Maarten van de Meent and Deyan P. Mihaylov and Serguei Ossokine and Michael Pürrer and Antoni Ramos-Buades and Ajit Kumar Mehta and Roberto Cotesta and Sylvain Marsat and Michael Boyle and Lawrence E. Kidder and Harald P. Pfeiffer and Mark A. Scheel and Hannes R. Rüter and Nils Vu and Reetika Dudi and Sizheng Ma and Keefe Mitman and Denyz Melchor and Sierra Thomas and Jennifer Sanchez},
      year={2023},
      eprint={2303.18039},
      archivePrefix={arXiv},
      primaryClass={gr-qc},
      url={https://arxiv.org/abs/2303.18039}, 
}

@article{Field_2014,
   title={Fast Prediction and Evaluation of Gravitational Waveforms Using Surrogate Models},
   volume={4},
   ISSN={2160-3308},
   url={http://dx.doi.org/10.1103/PhysRevX.4.031006},
   DOI={10.1103/physrevx.4.031006},
   number={3},
   journal={Physical Review X},
   publisher={American Physical Society (APS)},
   author={Field, Scott E. and Galley, Chad R. and Hesthaven, Jan S. and Kaye, Jason and Tiglio, Manuel},
   year={2014},
   month=jul }

@article{Blackman_2015,
   title={Fast and Accurate Prediction of Numerical Relativity Waveforms from Binary Black Hole Coalescences Using Surrogate Models},
   volume={115},
   ISSN={1079-7114},
   url={http://dx.doi.org/10.1103/PhysRevLett.115.121102},
   DOI={10.1103/physrevlett.115.121102},
   number={12},
   journal={Physical Review Letters},
   publisher={American Physical Society (APS)},
   author={Blackman, Jonathan and Field, Scott E. and Galley, Chad R. and Szilágyi, Béla and Scheel, Mark A. and Tiglio, Manuel and Hemberger, Daniel A.},
   year={2015},
   month=sep }

@article{Yoo_2022,
   title={Targeted large mass ratio numerical relativity surrogate waveform model for GW190814},
   volume={106},
   ISSN={2470-0029},
   url={http://dx.doi.org/10.1103/PhysRevD.106.044001},
   DOI={10.1103/physrevd.106.044001},
   number={4},
   journal={Physical Review D},
   publisher={American Physical Society (APS)},
   author={Yoo, Jooheon and Varma, Vijay and Giesler, Matthew and Scheel, Mark A. and Haster, Carl-Johan and Pfeiffer, Harald P. and Kidder, Lawrence E. and Boyle, Michael},
   year={2022},
   month=aug }

\end{document}